\documentclass[pdflatex]{sn-jnl}

\usepackage{graphicx}%
\usepackage{multirow}%
\usepackage{amsmath,amssymb,amsfonts}%
\usepackage{amsthm}%
\usepackage{mathrsfs}%
\usepackage[title]{appendix}%
\usepackage{xcolor}%
\usepackage{textcomp}%
\usepackage{manyfoot}%
\usepackage{booktabs}%
\usepackage{algorithm}%
\usepackage{algorithmicx}%
\usepackage{algpseudocode}%
\usepackage{listings}%
\usepackage{bm}%
\usepackage[authoryear]{natbib}%

\begin{document}

\title{A novel vortex generator for enhancing bileaflet mechanical heart valve performance proposed through detailed hemodynamic and data-driven dynamic mode decomposition studies}

\author{\fnm{A.} \sur{Chauhan}}

\author*[ ]{\fnm{C.} \sur{Sasmal}}
\email{csasmal@iitrpr.ac.in}

\affil{\orgdiv{Department of Chemical Engineering}, \orgname{Indian Institute of Technology Ropar}, \orgaddress{\city{Rupnagar}, \postcode{140001}, \state{Punjab}, \country{India}}}

\abstract{Prosthetic mechanical heart valves (MHVs), particularly the bileaflet types, have been one of the most advanced and commonly implanted heart valves for more than four decades. These bileaflet mechanical heart valves (BMHVs) offer superior durability and improved hemodynamics; however, they have certain drawbacks, including the need for lifelong anticoagulation to prevent blood clot formation and the associated risks of thromboembolism. To overcome these limitations, comprehensive research studies are deployed, among which controlling the overall hemodynamics by means of passive interventions, such as placing (multiple) vortex generators (VGs) on the heart valve leaflets, is one possibility. These VGs, in particular, are known to delay the separation of boundary layers formed around the leaflet surfaces, provided they are precisely positioned and have their chosen design features. Limited research has shown that VGs improve hemodynamics beyond a BHMV when rectangular-type VGs (RVGs) are considered. Specifically, configurations such as co-rotating and counter-rotating are investigated at an incidence angle of $23^{\circ}$; however, their location on the heart valve leaflets and their shape can prove vital, an aspect that has not been explored so far. Suffice it to add, the alteration in hemodynamic performance due to these VGs may often lead to an impact on several critical clinical indices such as the pressure gradients, turbulent intensity, wall shear stress (WSS), blood damage (BD), etc., which typically establish the overall functional efficiency of an MHV. To explicitly bridge this gap, we perform detailed numerical simulations using elliptic-type VGs (EVGs) and RVGs on the valve surface at various locations along the leaflets' leading edges, and ultimately, propose a novel vortex generator that offers superior hemodynamic performance and improved clinical parameters. Systematic comparisons and discussions are presented using velocity magnitude contours, kymographs of the axial velocity field, turbulent kinetic energy production (TKEP), pressure variation and its gradient, maximum and surface-averaged WSS, BD, etc., to support the selection of this novel VG. Not only have hemodynamic analyses been performed, but data-driven dynamic mode decomposition (DMD) of the vorticity field has also been conducted behind this proposed vortex generator, revealing hidden coherent flow structures that underlie the flow field and are critical to the design of an MHV. Altogether, the information obtained from this investigation can be leveraged to design and develop the next-generation bileaflet mechanical heart valve with better hemodynamics and superior performance.}


\maketitle

\section{Introduction}\label{Introduction}

Cardiovascular diseases (CVDs) remain one of the primary drivers of global mortality, with valvular heart diseases (VHDs), including atherosclerosis, regurgitation in the mitral, aortic, tricuspid or pulmonary valve, congenital defect, rheumatic heart disease (RHD), mitral valve prolapse, etc., contributing to an ascending clinical and socioeconomic burden~\citep{berry2012lifetime, gaidai2023global}. For instance, the World Health Organization (WHO) reports that an estimated $19.8$ million people died from CVDs in 2022, which represents around $32\%$ of the total global mortality~\citep{WHO}. Among these, $85\%$ deaths were due to a heart attack. With a growing population, these VHDs are rising globally and have a substantial impact on millions of people, a trend expected to escalate in the coming years~\citep{coffey2016modern, coffey2021global}. From a pathophysiological perspective, these abnormalities often disrupt unidirectional blood flow and trigger adverse ventricular remodelling, necessitating surgical or transcatheter replacement when pharmacological strategies fail to restore valvular mechanobiology~\citep{ajmone2023valvular}. Consequently, over the years, the increasing demand for heart valve prostheses, driven primarily by ageing populations and improved diagnostics, highlights the critical need to develop next-generation prosthetic heart valves (PHVs) with superior hemocompatibility, long-term durability, and biomimetic hemodynamics~\citep{singh2023polymeric, evangelista2025chronological,Zakaria2017}. These advancements are crucial for mitigating persistent complications such as device-induced thrombosis, hemolysis, and structural valve degeneration (SVD) over time~\citep{pibarot2009prosthetic}.

As a remedy for VHDs, surgical valve replacement typically employs two types of PHVs: mechanical or bioprosthetic heart valves, specifically designed to restore the natural features of a native heart valve, including long-term durability, smooth hemodynamics, thromboresistance, and implantability~\citep{head2017mechanical}. Although bioprosthetic heart valves (BHVs) feature superior initial thromboresistance, they suffer from SVD, limiting their operational lifespan to $10 - 15$ years~\citep{squiers2023structural}. On the other hand, the mechanical heart valve (MHV) exhibits exceptional structural longevity, exceeding 25 years, and therefore becomes a preferred choice for patients younger than 60 years of age~\citep{jaffer2016mechanical}. Based on their structural design features, these prosthetic MHVs are further categorised into three types, namely ball-and-cage, tilting disk, and bileaflet. Among these, the most widely used modern design for prosthetic MHV even today is the bileaflet mechanical heart valve (BMHV)~\citep{dewall2000evolution, gott2003mechanical}. However, certain clinical drawbacks of BMHVs persist. These include low thromboresistance due to platelet activation, predominantly caused by non-physiological blood flow patterns in the vicinity of the heart valve region, inducing hemolysis or clumping of blood; elevated wall shear stress (WSS); flow anomalies accelerating the shear-induced platelet activation; flow separation and recirculation; vortex shedding; generation of high turbulent kinetic energy (TKE), etc.~\citep{alemu2007flow, yoganathan2004fluid, dasi2009fluid, bluestein2000vortex}. 

Owing to these imperfections in the BMHVs, several researchers and scientists have proposed various design features to incorporate into existing prosthetic MHVs. This includes the meticulous investigation of hemodynamics in the vicinity of the heart valve leaflets, while accounting for accurate physiological conditions to minimise non-physiological flow phenomena. For instance, recent research has focused on surface modifications of heart valve leaflets via coatings or micro-nano texturing, resulting in the development of superhydrophobic (SH) surfaces. The main goal is to reduce the risk of blood clotting, primarily by minimising blood cell adhesion to the heart valve surface, thereby showcasing improved hemocompatibility and promoting the long-term durability of MHVs~\citep{yousefi2024surface}. However, despite the critical role of hemodynamics in determining the functional performance of prosthetic MHVs, only a limited number of investigations have examined how these SH surface modifications influence the associated flow dynamics. To name a few, Bark et al.~\citep{bark2017hemodynamic} and Hatoum et al.~\citep{hatoum2020impact} were the ones to report two contradicting outcomes, with the former study claiming that the hemodynamic parameters of clinical importance such as pressure drop and vorticity intensity were reduced on implementing SH coating compared to without it, whereas, the latter study revealed that the application of a SH surface coating on a 3D-printed BMHV led to an increase in both Reynolds shear stresses (RSS) and viscous shear stresses compared to the uncoated valve. Therefore, to shed light on these conflicting findings, our recent numerical simulations~\citep{chauhan2026impact} provided definitive results regarding the implementation of SH coatings on the surface of a BMHV. In particular, a partial and/or free-slip boundary condition was imposed on the leaflets of the BMHV, which typically arises from these SH coatings, and the system was tested under physiologically realistic pulsatile flow conditions at different time instants. Although the pressure drop across the valve region remains almost the same, the maximum WSS values increased ($> 21\%$ across all time instants) and the blood damage (BD) values reduced to about $8\%$ (during the mid-acceleration and peak systolic stage of the cardiac cycle) in the presence of free-slip condition.

In addition to introducing SH coatings on prosthetic MHVs, a limited number of studies have also incorporated vortex generators (VGs) with specific arrangements on the surface of the heart valve leaflets, aimed at the design and development of next-generation surface-modified MHVs. In particular, vortex generators are small passive flow-control devices designed to manipulate the local flow field by generating streamwise vortices that enhance momentum exchange between the near-wall and core flow regions~\citep{jayanarasimhan2025overview}. This improved mixing helps delay flow separation, reduce recirculation zones, and promote a more uniform velocity distribution. As a result, vortex generators are widely used in aerospace and turbomachinery applications to improve aerodynamic or hydrodynamic performance, reduce pressure losses, and optimise overall system efficiency with minimal additional energy input~\citep{zhao2022researches,jirasek2005vortex}. The use of VGs in biomedical applications, particularly for mechanical heart valves, has also been explored. For instance, using the particle image velocimetry (PIV) technique for flow field visualisation, Dasi et al.~\citep{dasi2008passive} investigated the hemodynamics of a BMHV by mounting rectangular and hemispherical vortex generators on the downstream surfaces of it, adjacent to the b-datum leaflet edge. They found that the VGs spatially disperse and dissipate the coherent leakage jet structure emanating from the b-datum line, resulting in a significant diminution of turbulence stresses, particularly with the rectangular VG configuration. Subsequently, from the same research group, Murphy et al.~\citep{murphy2010reduction} performed \textit{in-vitro} experiments using a steady model of the transient b-datum line jet with and without the application of VGs. They found that the presence of VGs ultimately reduced the blood's propensity for thrombus formation, likely due to a $10-20\%$ reduction in peak turbulent shear stress (TSS) in the jet. In another experimental study by Hatoum and Dasi~\citep{hatoum2019reduction}, various arrangements of rectangular VGs mounted downstream of the BMHV were investigated. In particular, high-resolution PIV was employed under physiological pulsatile inflow conditions for four sets of VG arrangements, namely, four equally spaced co-rotating VGs, eight equally spaced counter-rotating VGs, four far-spaced counter-rotating VGs, and four closely spaced counter-rotating VGs. Overall, i.e., with or without VGs, the co-rotating VG setup reduced RSS and improved the pressure gradient. Moreover, the co-rotating VGs led to a delayed flow separation and a more homogenised, streamlined flow transition compared with any counter-rotating VG configuration. From the perspective of computational fluid dynamics (CFD), Wang et al.~\citep{wang2022controlling} carried out numerical simulations to scrutinise the effectiveness of VGs on BMHV. Co-rotating and counter-rotating VG arrangements were tested in conjunction with the absence of any VGs. Their results revealed that VGs in a co-rotating configuration improve hemodynamics compared to the counter-rotating case, as reflected in RSS, TKE production, velocity distribution, etc., especially in the vicinity of the heart valve. Compared with the absence of VG, the fraction of BD was reported to be $4.7\%$ lower and $3.7\%$ higher in the co-rotating and counter-rotating setups, respectively. Nitti et al.~\citep{nitti2022numerical} conducted numerical simulations for three configurations of MHVs: St. Jude medical valve (SJMV), Lapeyre-Triflo FURTIVA valve (LTFV) with three leaflets, and a SJMV with VGs (SJMVVG). Their study revealed that VGs introduce instabilities that disrupt the Kármán-like vortex shedding downstream of the heart valve leaflets, thereby reducing TKE intensity at peak flow and lowering local RSS.

Therefore, the literature cited herein demonstrates the effectiveness of integrating rectangular-type VGs onto the surface of BMHV in improving overall hemodynamics. However, it should be noted that the analyses used to establish the superiority of VGs were questionable and, in some cases, did not reflect actual physiological conditions. For instance, Dasi et al.~\citep{dasi2008passive} analysed the results on a two-dimensional plane to demonstrate the efficiency of VGs, whereas the flow is inherently three-dimensional in flows past an MHV. This is, in fact, mentioned by Dasi et al. in their study as one of the limitations in assessing the role of VGs in improving flow-structure. Furthermore, the numerical study by Wang et al.~\citep{wang2022controlling} employed a steady inflow condition; however, the flow in a heart valve is inherently pulsatile, which they also identified as a drawback of the study. Therefore, the role of VGs in improving the performance of a BMHV is still questionable to a certain extent. The present study aims to reduce that gap in understanding by performing full-scale three-dimensional numerical simulations that account for physiologically realistic pulsatile flow conditions across various VG configurations, some of which have already been investigated and others proposed herein. Ultimately, we develop a novel vortex generator design through detailed hemodynamic studies and analyses, including flow field analysis, turbulent energy production, pressure distribution, and clinical parameters such as wall shear stress. We propose this design not only from the perspective of hemodynamic analyses but also from that of dynamic mode decomposition (DMD). It is a powerful data-driven modal decomposition technique that extracts the dominant spatiotemporal flow structures and their associated frequencies directly from time-resolved velocity or vorticity fields~\citep{schmid2010dynamic,schmid2011application,schmid2022dynamic}. This technique has recently been utilised extensively to understand cardiac flow dynamics and to analyse heart disease~\citep{Groun2022}. Particularly in flow-through bileaflet mechanical heart valves, DMD could provide valuable insights into the evolution of coherent vortical structures, leakage jets, shear layers, and turbulent fluctuations, which are closely associated with blood damage and platelet activation. Unlike conventional statistical methods, DMD can identify dynamically important flow modes that drive flow instability and energy transport, enabling a detailed understanding of the mechanisms governing adverse hemodynamics. Consequently, DMD serves as an effective tool for evaluating and optimising heart valve designs, assessing passive flow-control strategies such as vortex generators, and developing next-generation prosthetic valves with improved hemodynamic performance.

\section{Problem formulation and governing equations}

\begin{figure}
    \centering
    \includegraphics[trim=0cm 0cm 0cm 0cm,clip,width=13cm]{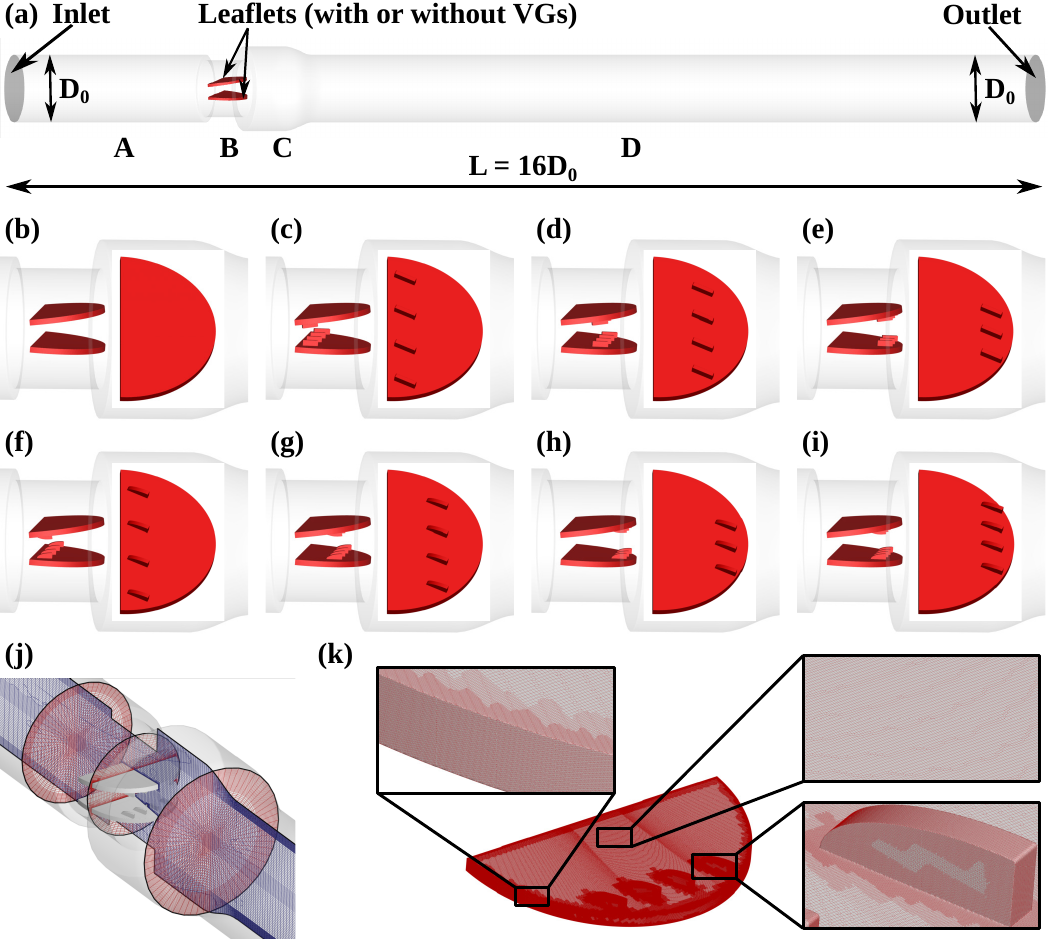}
    \caption{(a) Present cardiovascular domain, which consists of four regions: \textbf{A}- ventricular side chamber, \textbf{B}- valve, \textbf{C}- sinus expansion and \textbf{D}- aortic side chamber. The zoomed view near the valve region for (b) standard SJMV, (c) SJMVRVG-I, (d) SJMVRVG-II, (e) SJMVRVG-III, (f) SJMVEVG-I, (g) SJMVEVG-II, (h) SJMVEVG-III, and (i) SJMVEVG-IV. For ease of visualisation, the leaflet surface on which VGs are placed is shown in the corresponding sub-Figures. (j) An example of the grid structure used is shown for the SJMVEVG-IV configuration with explicit refinement near the valve leaflets and artery walls to capture the steep gradients of primitive variables. (k) The zoom-in distribution of the polyhedral-type grid cells generated on the surface of heart valve leaflets. The other information for different VG configurations is presented in Table~\ref{table:HV_design}.} 
    \label{Geometry}
\end{figure}

\begin{figure}
    \centering
    \includegraphics[trim=0cm 0cm 0cm 0cm,clip,width=8cm]{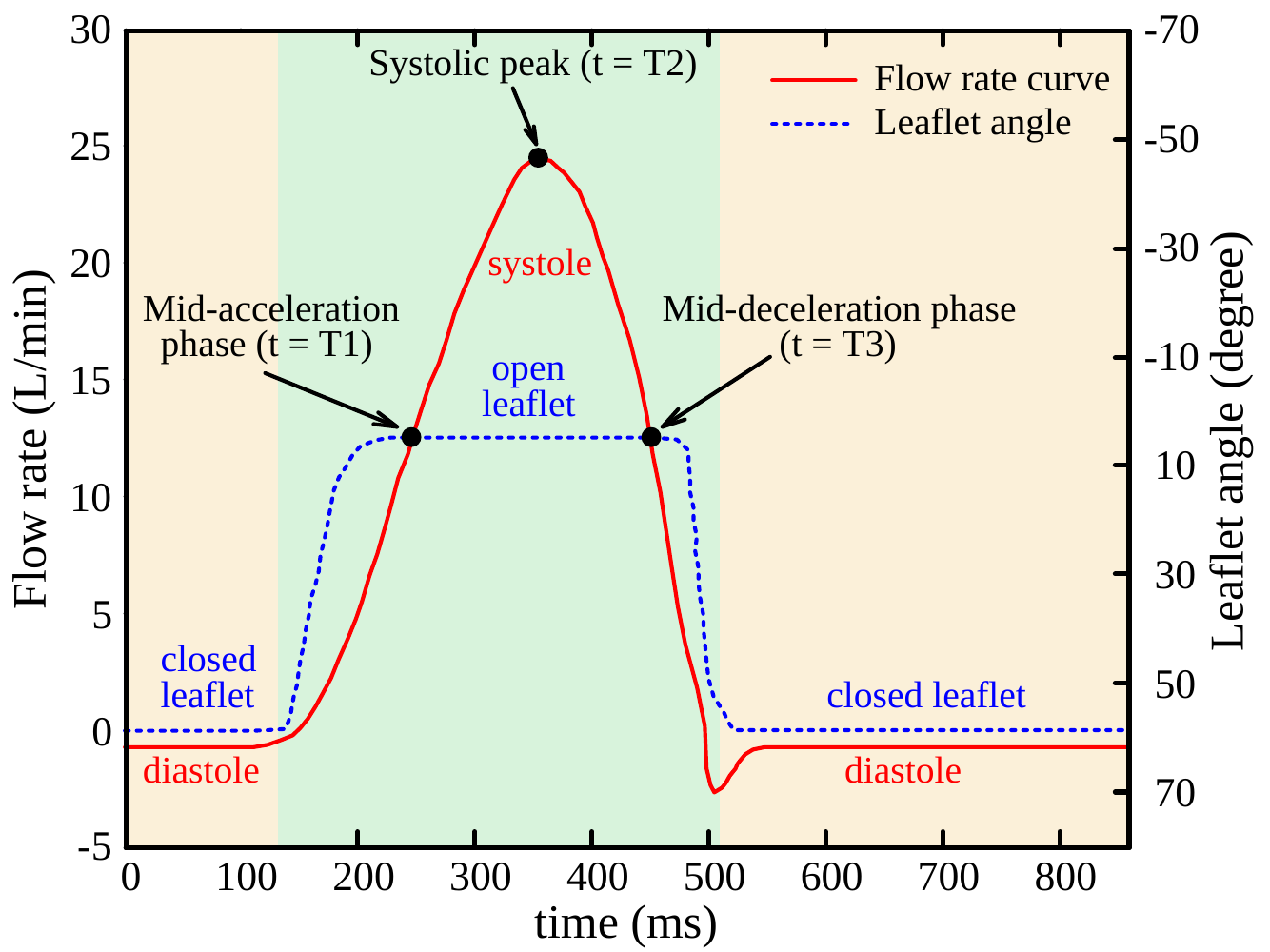}
    \caption{Variation of the pulsating flow rate signal (primary axis) and the associated leaflet angle (secondary axis) over one cardiac cycle of 860 ms, dictating a heart rate of 70 beats/min~\citep{Yun2014}. Key timestamps used in the present study (T1, T2, and T3) for discussing the results are shown in the same figure.} 
    \label{flowCurve}
\end{figure}

Considering the standard St. Jude medical valve (SJMV) of 23 mm diameter used widely in clinical heart valve replacements, we construct a cardiovascular domain as shown in sub-Fig.~\ref{Geometry}(a). Such a valve model has been widely adopted in both experimental and numerical studies~\citep{Ge2005, Yun2014}. In this study, we place the various configurations of vortex generators (VGs) on the heart valve leaflets, which are shown in sub-Figs.~\ref{Geometry}(b)-(i). The other details, such as the distance of VGs from the leading edge of the leaflets, their spacing, number, type (rectangular (RVG) or elliptic (EVG)), etc., are listed in Table~\ref{table:HV_design}. For all cases with VGs in this study, the length, width, and height of the VG are set to 2.8 mm, 1 mm, and 1 mm, respectively, for the co-rotating configuration at an incidence angle of $23^{\circ}$. A similar arrangement with RVGs, specifically the SJMVRVG-I configuration, has also been adopted in the previous studies~\citep{hatoum2019reduction, wang2022controlling}. Furthermore, the power-law model that captures the shear-thinning behaviour of blood is employed, as detailed in a series of our recent works~\citep{chauhan2024influence,chauhan2024hemodynamics,chauhan2026impact}. The pulsating nature of blood is imposed over the cardiac cycle as illustrated in Fig.~\ref{flowCurve}. It is to be noted that the analysis performed in the present numerical framework neglects the hinge mechanism of the heart valve leaflets. In particular, our focus is on the systolic time interval when the valve is fully open, during which both leaflets of the standard SJMV remain stationary at an angle of $5^\circ$ with respect to the xz-plane. A similar strategy has also been employed in previous experimental and numerical studies to elucidate the hemodynamics past a BMHV~\citep{bark2017hemodynamic, hatoum2020impact, wang2022controlling, Ge2005}. For the unsteady numerical simulations carried out in the present study, the following equations are solved in their dimensional form, assuming incompressible blood flow. 
\begin{table}
  \caption{Details related to the various configurations of VGs placed on the heart valve leaflets.}
  \label{table:HV_design}
  \centering
  \begin{tabular}{ccccc}
  \toprule
  \multirow{2}{*}{\textbf{Configuration}} & \multirow{2}{*}{\textbf{Type of VG}} & \textbf{Number of VGs} & \textbf{Distance from} & \textbf{Spacing between} \\
  &  & \textbf{on each leaflet} & \textbf{leading edge [mm]} & \textbf{consecutive VG [mm]} \\
  \midrule
  SJMV & $-$ & $-$ & $-$ & $-$ \\
  SJMVRVG-I & Rectangular & 4 & 1 & 5 \\
  SJMVRVG-II & Rectangular & 4 & 5 & 4 \\
  SJMVRVG-III & Rectangular & 3 & 8 & 3 \\
  SJMVEVG-I & Elliptic & 4 & 1 & 5 \\
  SJMVEVG-II & Elliptic & 4 & 5 & 4 \\
  SJMVEVG-III & Elliptic & 3 & 8 & 3 \\
  SJMVEVG-IV & Elliptic & 4 & 7 & 3 \\
  \bottomrule
  \end{tabular}
  \end{table}
\\
Continuity equation:
\begin{equation} \label{eq:continuity}
 \bm{\nabla} \cdot \bm{u} = 0
\end{equation}
Momentum equation:
\begin{equation} \label{eq:momentum}
    \rho \left( \frac{\partial \bm{u}}{\partial t} + \bm{u} \cdot \bm{\nabla u} \right) = - \bm{\nabla} p + \bm{\nabla} \cdot \bm{\tau} 
\end{equation}
Where, $\bm{u}$ is the velocity vector, $\bm{\nabla}$ is the gradient operator, $t$ is the time, $p$ is the pressure, $\bm{\tau}$ is the extra-stress tensor, and $\rho$ is the density of blood (taken as $1060$ $kg/m^{3}$). The extra-stress tensor, $\bm{\tau}$, is calculated using the following constitutive relation~\citep{Chhabra2011}:
\begin{equation} \label{eq:constitutive}
   \bm{\tau} = {\eta}~\bm{{\dot {\gamma}}} 
\end{equation}
Where, $\bm{{\dot {\gamma}}} = \frac{1}{2} \left(\bm{\nabla u} + \bm{\nabla u^T}\right)$ is the strain-rate tensor and $\eta = \eta\left(|\bm{\dot{\gamma}}|\right)$ represents the apparent shear viscosity of blood. As mentioned before, in the present study, we have adopted the power-law rheological model to calculate the apparent viscosity of the blood, as follows:
\begin{equation} \label{eq:power-law}
   {\eta} = k {|\bm{\dot{\gamma}}|}^{n-1}, \qquad {\eta_0} \leq {\eta} \leq {\eta_{\infty}} 
\end{equation}
Where $k$, $n$, $\eta_0$, and $\eta_{\infty}$ are blood consistency coefficient, flow behaviour index, zero-shear viscosity, and infinite (or high) shear viscosity, respectively. It is worth pointing out here that Eq.~\eqref{eq:power-law} predicts an infinite and a zero value of the apparent shear viscosity in the limits of low and high values of $|\bm{\dot{\gamma}}|$, respectively. Such predictions are not physically realistic and represent a fundamental drawback of this non-Newtonian power-law model~\citep{Chhabra2011}. To address this limitation in the present study, we adopt a cut-off strategy. In particular, for low values of $|\bm{\dot{\gamma}}|$, the apparent viscosity is bounded as $\eta = \eta{\left(|\bm{\dot{\gamma}}| \rightarrow 0\right)} = \eta_0 = 0.0548$ Pa·s, and for high values of $|\bm{\dot{\gamma}}|$, it is limited as $\eta = \eta{\left(|\bm{\dot{\gamma}}| \rightarrow \infty\right)} = \eta_{\infty} = 0.0035$ Pa·s~\citep{chauhan2026impact}.

\section{Numerical details}\label{NumDetail}
In this study, we deploy the open-source computational fluid dynamics (CFD) toolbox OpenFOAM (version 7)~\citep{openfoam} to numerically solve the governing equations discussed in the previous section such as the continuity equation (Eq.~\eqref{eq:continuity}), the momentum equation (Eq.~\eqref{eq:momentum}), and the power-law constitutive equation (Eq.~\eqref{eq:power-law}). Among the available choices of solvers in OpenFOAM, we utilise the transient \textit{pisoFoam} solver that incorporates pressure-velocity coupling through the Pressure Implicit with Splitting of Operators (PISO) method~\citep{ferziger2002computational}. This method, in particular, is effective for simulating unsteady flow problems and has been shown to be more effective than other methods, such as the Semi-Implicit Method for Pressure-Linked Equations (SIMPLE). A detailed discussion of the various discretisation techniques employed in the present study is available in our recent works~\citep{chauhan2024influence, chauhan2024hemodynamics}; for brevity, we do not discuss them in detail here. Moreover, in the present study, we run the simulations for three cardiac cycles (with one cycle lasting until $t = 860$ ms) using 400 cores in parallel via the message-passing interface (MPI) available within the OpenFOAM framework. Each simulation required approximately $72 - 84$ hours of wall-clock time, totalling about $28,000 - 34,000$ total computational hours, on a supercomputing cluster equipped with Intel Xeon Platinum 8268 (2.9 GHz) processors and 960 GB of RAM per node.

Furthermore, we employ the following boundary conditions on different boundaries of the present cardiovascular system: At the inlet patch, a cardiac pulse is used for velocity, as detailed in our earlier works~\citep{chauhan2024influence, chauhan2024hemodynamics, chauhan2026impact}, whereas a zero-gradient condition is prescribed for pressure. At the outlet patch, a Neumann-type boundary condition and a constant value of 100 mmHg are utilised for velocity and pressure, respectively. On the surface of the artery walls and the surface of the leaflets (with or without VGs), a no-slip condition was imposed for velocity and a zero-gradient condition for pressure.

It is worth noting that a meticulous investment in selecting optimal grid and time-step sizes was already made in our previous study through grid and time-step convergence studies~\citep{chauhan2024influence}; therefore, this exercise has not been repeated here. However, for completeness, some noteworthy aspects, including those related to generating grid cells in the cardiovascular domain, are mentioned here. For instance, the preliminary grid, consisting of hexahedral cells (without the leaflets), was first created using the \textit{blockMesh} subroutine in OpenFOAM. Thereafter, we make use of the robust \textit{snappyHexMesh} utility that constructs the complex leaflet structures (both with and without the VGs) present in the valve region, i.e., region B in sub-Figure~\ref{Geometry}, with the grid cells typically refined close to the surface of these solid walls. This local refinement near the leaflet structures in the present study is achieved via polyhedral cells, which are crucial for capturing steep gradients in velocity, stress, pressure, etc., with high accuracy in the present numerical computations. Not only that, the presence of polyhedral cells is well known to strengthen the overall grid quality and boost the numerical stability of the currently implemented code, owing to their attenuated spatial tuning to refinements in the vicinity of leaflet structures and artery walls. An example of the grid structure used in the present computations is shown in sub-Figs.~\ref{Geometry}(j) and (k) for the SJMVEVG-IV configuration. Additionally, a thorough validation of the present implemented numerical code has already been reported in our previous studies~\citep{chauhan2024influence, chauhan2024hemodynamics} for the \textit{in-vitro} experiments and \textit{in-silico} numerical results of Yun et al.~\citep{Yun2014}, wherein a similar cardiovascular system has been used incorporating a 23 mm diameter SJMV. We, therefore, have not repeated this exercise for conciseness in the present study and discuss the new results obtained in the subsequent sections.

\section{Results and discussion}

\begin{figure}
    \centering
    \includegraphics[trim=0cm 0cm 0cm 0cm,clip,width=13cm]{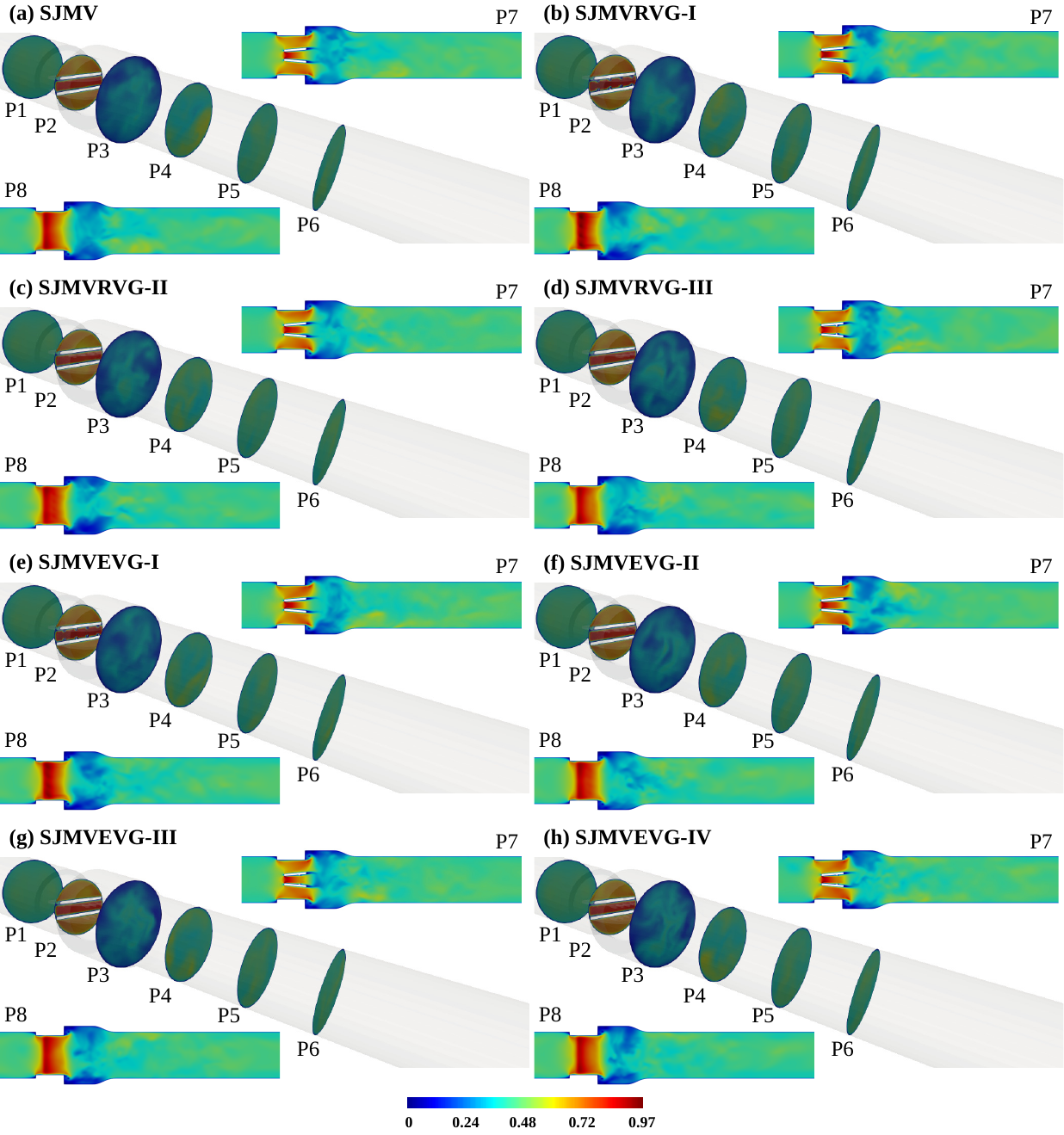}
    \caption{Contours of velocity magnitude (in units of m/s) across distinct cross-sectional planes (P1--P8) for various configurations presented in Table~\ref{table:HV_design} during $t = \text{T1} \approx 1965$ ms (mid-acceleration phase).} 
    \label{Umagt1}
\end{figure}

As mentioned in the previous section, the numerical simulations conducted in this study are of similar settings to those reported in our earlier work~\citep{chauhan2024influence}, i.e., three cardiac cycles, with results primarily discussed and analysed for the final periodic cycle during mid-acceleration (T1), peak systolic (T2), and mid-deceleration (T3) phases. Additionally, at certain instances, time-averaged results over the last cardiac cycle have also been presented to evaluate parameters such as turbulent kinetic energy production (TKEP) and blood damage (BD). Other hemodynamic metrics evaluated here include velocity magnitude contours, the kymograph of the axial velocity component, pressure variation and its gradient along the centerline, maximum and surface-averaged WSS, among others, for various VG combinations.

\subsection{Surface distribution of velocity field}

We begin with the variation of velocity magnitude contours, as showcased in Fig.~\ref{Umagt1} at eight different cross-sectional planes (P1-P8) during the mid-acceleration phase ($t = \text{T1} \approx 1965$ ms) of the cardiac cycle for various configurations presented in Table~\ref{table:HV_design}. These eight planes are selected such that the normal vector of P1--P6 lies along the x-direction, where P1 is located in region A (upstream of the valve), P2 is situated in region B (valve region), P3 is placed in region C (sinus region) and P4--P6 are stationed in region D (downstream of the valve and sinus region). The other two planes, namely P7 and P8, are positioned in a manner that the normal vector aligns with the z- and y-directions, respectively. From Fig.~\ref{Umagt1} it can be evidenced that a relatively uniform velocity field is seen in the central regions of the ventricular and aortic chambers for the SJMV case (see planes P1, P5 and P6 in sub-Fig.~\ref{Umagt1}(a)). This trend remains consistent for the cases with the incorporation of both rectangular and elliptic type VGs, sub-Figs.~\ref{Umagt1}(b)--(h). In contrast, the non-uniformity in the velocity magnitude is explicitly seen in the region C (sinus region), irrespective of the case, i.e., with or without VGs, due to the flow separation near the trailing edges of the leaflets, leading to the origination of low-velocity vortical structures (plane P3). The three distinct jets (one central and two lateral) formed in the SJMV case upon trifurcation are also notable in the presence of VGs on plane P7; however, the magnitude, length and/or width of the high magnitude central jet is imperceptibly altered based on the location and choice of VG. For instance, the length and magnitude of velocity in the central region are more in the presence of VGs (sub-Figs.~\ref{Umagt1}(b)--(h)) than that observed for the SJMV case (sub-Fig.~\ref{Umagt1}(a)). This is because the central area decreases with the additional incorporation of VGs, thereby increasing the velocity magnitude to satisfy the mass conservation principle. Not only this, the high velocity zone visible particularly in the valve region on plane P8 shows regions of high velocity magnitude for SJMVRVG-I (sub-Fig.~\ref{Umagt1}(b)) and SJMVEVG-I (sub-Fig.~\ref{Umagt1}(e)) compared to SJMV (sub-Fig.~\ref{Umagt1}(a)). Moreover, the velocity magnitude zone near the valve region retains its concave shape for most of the cases, similar to the SJMV except for the SJMVRVG-III (sub-Fig.~\ref{Umagt1}(d)) and SJMVEVG-III (sub-Fig.~\ref{Umagt1}(g)) case, wherein the non-uniformity occurs due to VGs situated near the trailing edges of the leaflets. These observations clearly suggest the collective impact of incorporating rectangular and/or elliptic type VGs on the velocity profile distribution and the jet dynamics within the cardiovascular domain.

\begin{figure}
    \centering
    \includegraphics[trim=0cm 0cm 0cm 0cm,clip,width=13cm]{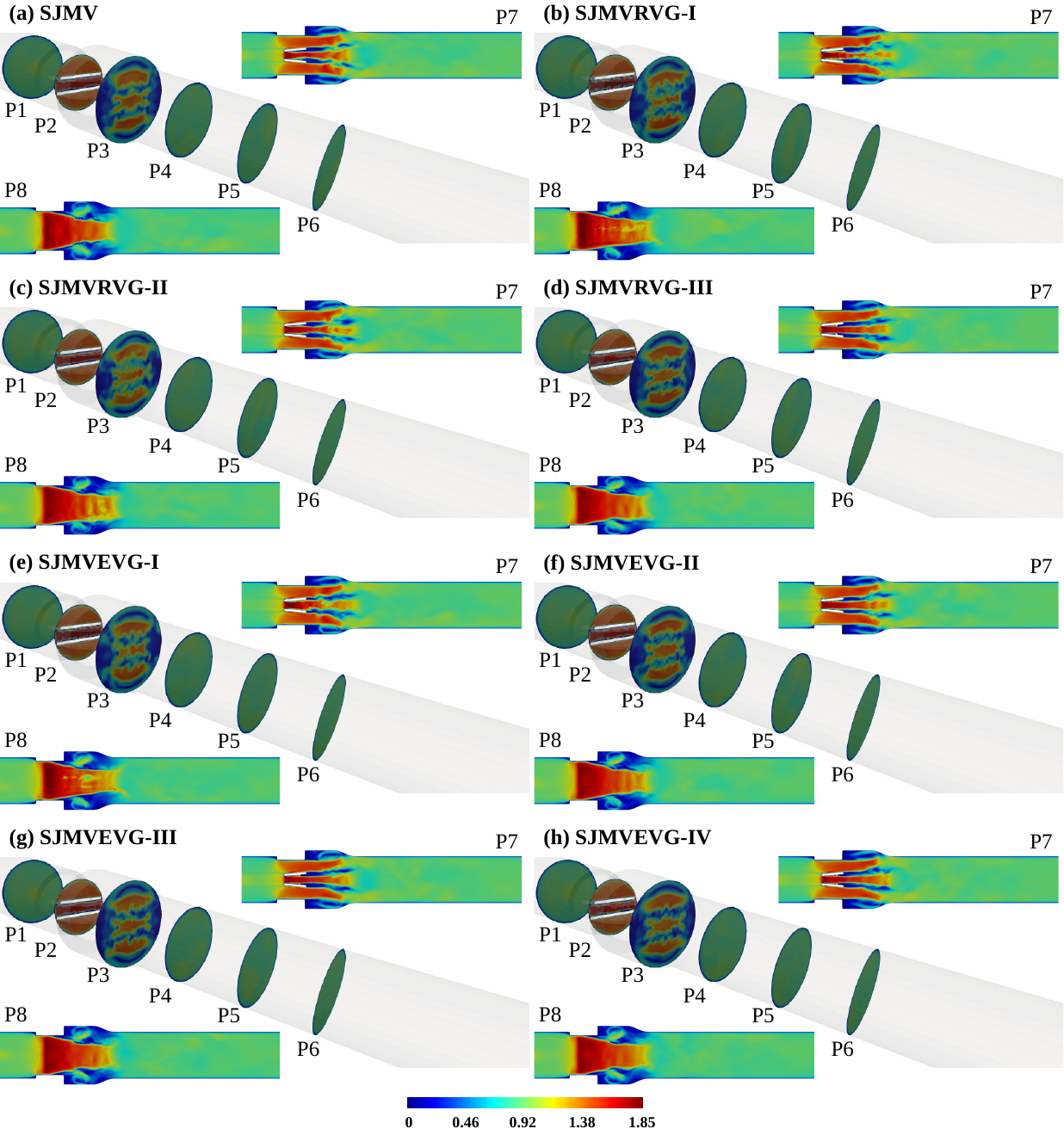}
    \caption{Contours of velocity magnitude (in units of m/s) across distinct cross-sectional planes (P1--P8) for various configurations presented in Table~\ref{table:HV_design} during $t = \text{T2} \approx 2074$ ms (peak systolic phase).} 
    \label{Umagt2}
\end{figure}

\begin{figure}
    \centering
    \includegraphics[trim=0cm 0cm 0cm 0cm,clip,width=13cm]{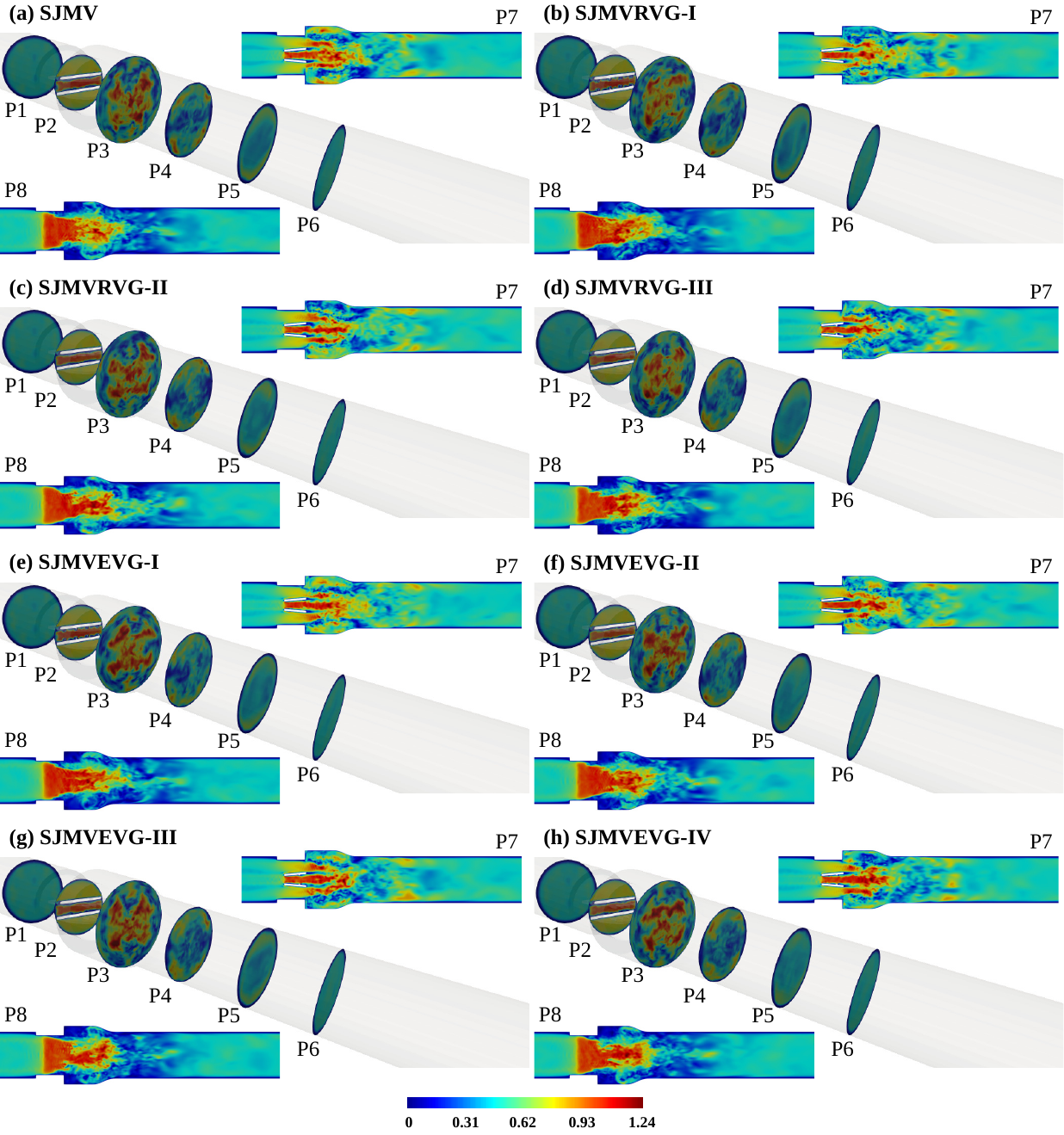}
    \caption{Contours of velocity magnitude (in units of m/s) across distinct cross-sectional planes (P1--P8) for various configurations presented in Table~\ref{table:HV_design} during $t = \text{T3} \approx 2170$ ms (mid-deceleration phase).} 
    \label{Umagt3}
\end{figure}

During the systolic peak of the cardiac cycle ($t = \text{T2} \approx 2074$ ms), the role of VGs is seen to be significantly greater than previously observed during the mid-acceleration phase, as demarcated in Fig.~\ref{Umagt2}. First, the width of the high magnitude lateral jets extending in the sinus region is seen to be less at the tip for the cases with VGs (sub-Figs.~\ref{Umagt2}(b)--(h)), as compared to without it, i.e., the SJMV case (sub-Fig.~\ref{Umagt2}(a)), evident in plane P7. Second, the length and breakage of the high-magnitude central jet are observed to be sensitive to the location and/or type of VGs placed on the valve leaflets. For instance, when the VGs are placed near the leading edges, the large central jet breaks into two or more parts, as noticed in sub-Fig.~\ref{Umagt2}(e) and sub-Fig.~\ref{Umagt2}(b) for elliptic (SJMVEVG-I) and rectangular (SJMVRVG-I) type VGs, respectively. Moreover, when the distance of the VGs is further increased from the leading edges (e.g., SJMVRVG-II and SJMVEVG-II), this breakage of the central jet occurs only for rectangular type VG (sub-Fig.~\ref{Umagt2}(c)), but not for elliptic type VG (sub-Fig.~\ref{Umagt2}(f)). Notably, when the VGs are placed near the trailing edges, the central jet does not break at all, as shown in the sub-Figs.~\ref{Umagt2}(d), (g) and (h). Not only this, the low velocity magnitude region downstream of the trailing edges of the two leaflets (between the central jet and lateral jet) increases in length without any breakage, unlike the SJMV case where breakage is seen (sub-Fig.~\ref{Umagt2}(a)). Due to the breakage of either a high magnitude central jet or a low magnitude jet appearing downstream the trailing leaflet edges in the case of VGs, the mushroom-shaped velocity field evident in plane P3 of sub-Fig.~\ref{Umagt2}(a) for SJMV also gets distorted. Altogether, these observations again highlight the intricate features of adopting VGs during the peak phase of the pulsating blood flow.

As the flow traverses to the mid-deceleration phase of the cardiac cycle ($t = \text{T3} \approx 2170$ ms), the addition of VGs to the SJMV case further increases the complexity of velocity patterns existing at this time instant due to the presence of an adverse pressure gradient in the cardiovascular system. As illustrated in Fig.~\ref{Umagt3}, the jet structures clearly become highly disrupted and irregular irrespective of the presence or absence of VGs. The escalation of chaotic behaviour is particularly pronounced in the sinus region and downstream on planes P3, P4, P7, and P8. The placement of VGs reduces the velocity field disruption for some configurations; for instance, see the case of SJMVEVG-IV (sub-Fig.~\ref{Umagt3}(h)), for which the leading front of the central jet seems to be less disrupted or broken into fragments compared to that seen for the case without VGs, i.e., SJMV. This can be seen both in planes P7 and P8 of sub-Fig.~\ref{Umagt3}(h). This could reduce damage to blood cells, which is evaluated and discussed later in this section of the present study using the blood damage index.   

\begin{figure}
    \centering
    \includegraphics[trim=0cm 0cm 0cm 0cm,clip,width=13cm]{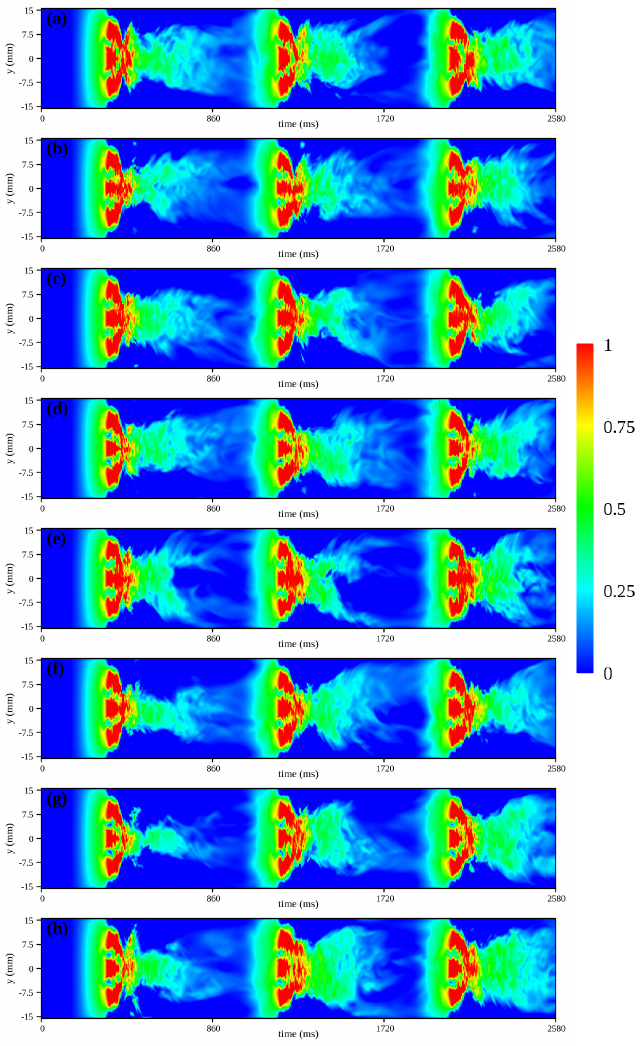}
    \caption{Kymograph for axial velocity profile (in units of m/s) at a line situated in the sinus region ($x=4.3D_0$, $z=0$) during three cardiac cycles: (a) SJMV, (b) SJMVRVG-I, (c) SJMVRVG-II, (d) SJMVRVG-III, (e) SJMVEVG-I, (f) SJMVEVG-II, (g) SJMVEVG-III, and (h) SJMVEVG-IV.} 
    \label{kymograph}
\end{figure}

To further comment on the temporal evolution of the flow profile, we track the kymographs of axial velocity along a probe line situated in the region proximal to the heart valve. Note that, for completeness and broader visualisation, we hereby picturize the kymographs in Fig.~\ref{kymograph} for all three cardiac cycles and for all the cases listed in Table~\ref {table:HV_design}. Clearly, before blood flow reaches its systolic peak, i.e., $t \lesssim 354$ ms, irrespective of the presence or absence of VGs, the kymograph patterns look almost the same, except in the middle region ($y = 0$), which is similar to the different noted magnitudes of central jets in Fig.~\ref{Umagt1} and Fig.~\ref{Umagt2}. However, as soon as blood starts to decelerate ($t > 350$ ms), changes in flow patterns are clearly evident due to the adverse pressure gradient in the presence of VGs, compared with their absence. For instance, temporal disruptions are reduced when VGs are present on the heart valve leaflets (see sub-Figs.~\ref{kymograph}(b)-(h)) compared to the standard SJMV case (sub-Fig.~\ref{kymograph}(a)). In particular, these disruptions are observed to further diminish when the distance of the VGs from the leading edge increases, and this amelioration is more evident in EVGs than in RVGs (see sub-Figs.~\ref{kymograph} (f) and (g) in comparison to sub-Figs.~\ref{kymograph} (c) and (d), respectively). Additionally, the relative magnitude of velocity slightly drops when either type of VG, i.e., rectangular or elliptic, is introduced. Akin to the trend seen for cycle-1 (upto $t = 860$ ms), the kymograph variation for the next two cycles (cycle-2: $860 < t \leq 1720$ ms and cycle-3: $1720 < t \leq 2580$ ms) is more or less similar, except for the deceleration stage, wherein the presence of an adverse pressure gradient during these time instants increases the complexity and/or irregularities in the velocity fields, as also observed before during the mid-deceleration stage in Fig.~\ref{Umagt3}. Overall, these kymographs suggest that momentum retention is better conserved when VGs are incorporated onto the leaflet surfaces, at least in the vicinity of the heart valve leaflets.

\subsection{Turbulent kinetic energy production}

\begin{figure}
    \centering
    \includegraphics[trim=0cm 0cm 0cm 0cm,clip,width=13cm]{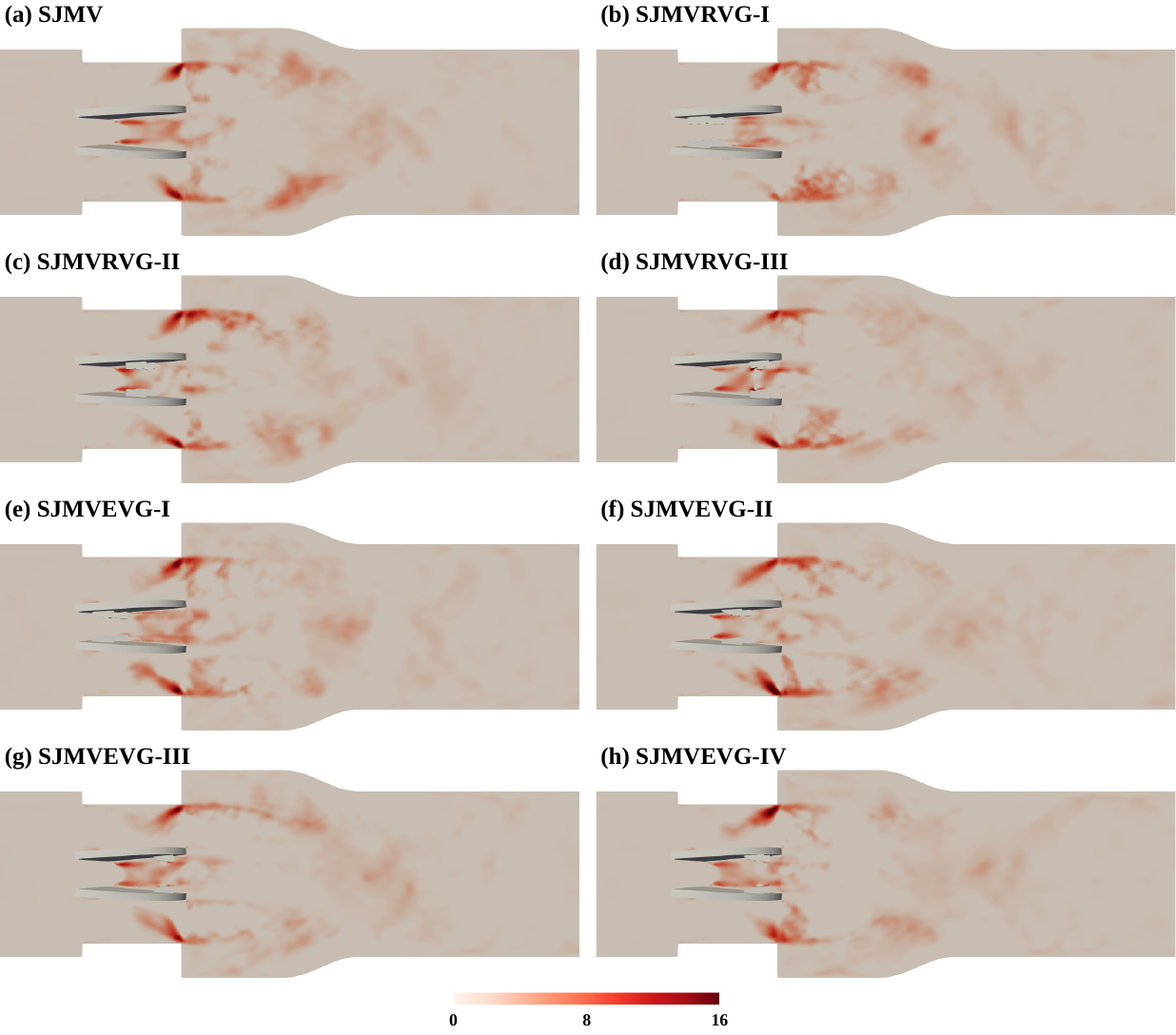}
    \caption{TKEP evaluated after the entire third cycle for various configurations presented in Table~\ref{table:HV_design} on $z = 0$ plane.} 
    \label{TKEP}
\end{figure}

\begin{figure}
    \centering
    \includegraphics[trim=0cm 0cm 0cm 0cm,clip,width=8cm]{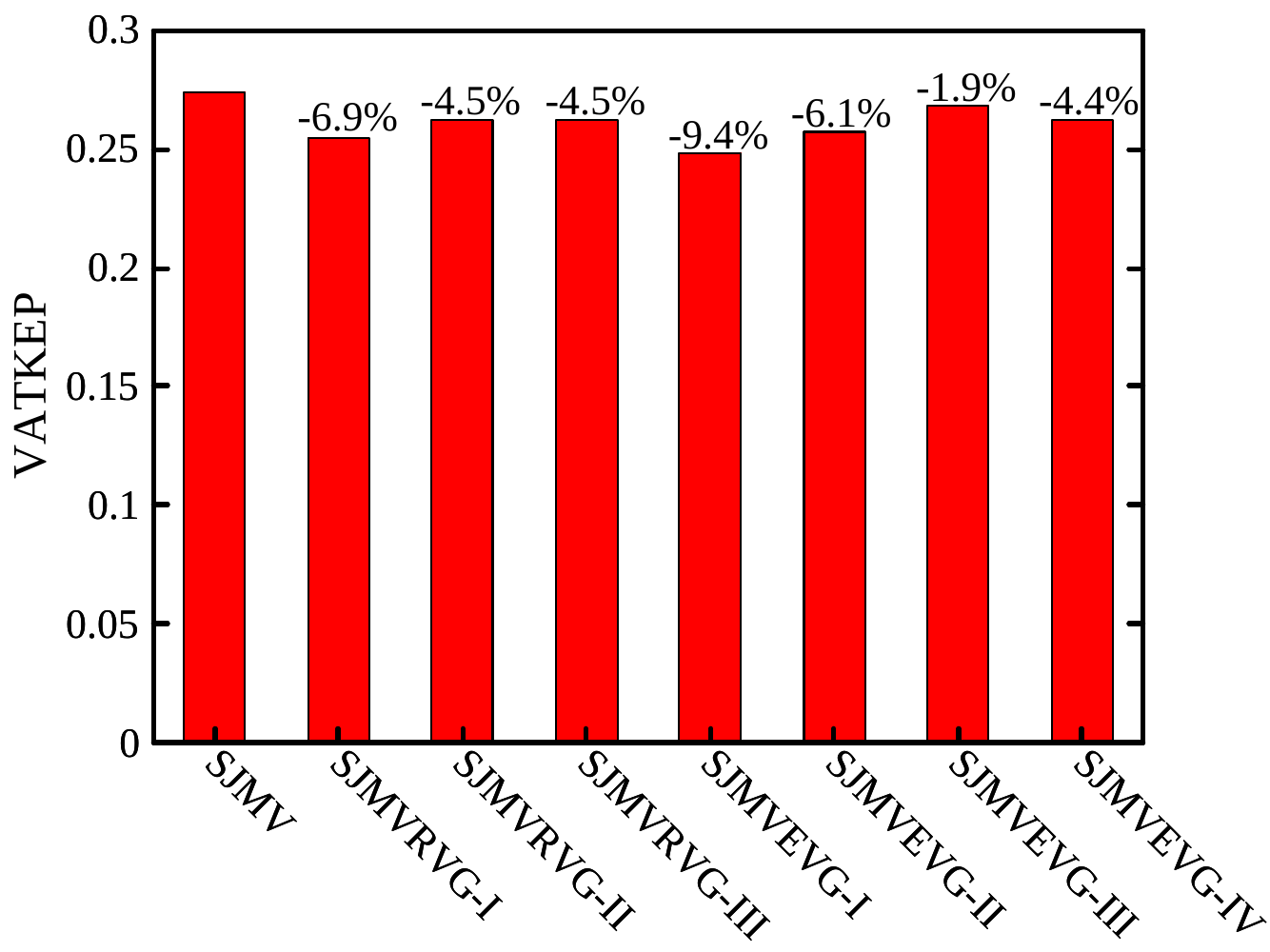}
    \caption{Histogram plot showing the VATKEP calculated in the valve and sinus region, i.e., region B and C, for various configurations presented in Table~\ref{table:HV_design}.} 
    \label{VATKEP}
\end{figure}

To further characterise the turbulent flow features in the vicinity of the heart valve, we calculate the turbulent kinetic energy production (TKEP) calculated using the following expression: 

\begin{multline} \label{eq:TKEP}
    \text{TKEP} = -\left[\left(\overline{u^{'}_x u^{'}_x} \times \frac{\partial \overline{u_x}}{\partial x} \right) + \left(\overline{u^{'}_y u^{'}_y} \times \frac{\partial \overline{u_y}}{\partial y} \right) + \left(\overline{u^{'}_z u^{'}_z} \times \frac{\partial \overline{u_z}}{\partial z} \right) \right. \\
    \left. + \left(\overline{u^{'}_x u^{'}_y} \times \left(\frac{\partial \overline{u_x}}{\partial y} + \frac{\partial \overline{u_y}}{\partial x}\right) \right) + \left(\overline{u^{'}_x u^{'}_z} \times \left(\frac{\partial \overline{u_x}}{\partial z} + \frac{\partial \overline{u_z}}{\partial x}\right) \right) \right. \\ 
    \left. + \left(\overline{u^{'}_y u^{'}_z} \times \left(\frac{\partial \overline{u_y}}{\partial z} + \frac{\partial \overline{u_z}}{\partial y}\right) \right) \right] 
\end{multline}

Where $\overline{u^{'}_i u^{'}_i}$ and $\overline{u^{'}_i u^{'}_j}$ ($i \neq j$) are the Reynolds normal and shear stress, respectively, calculated after the completion of the entire third cycle. Also, $\overline{u_i}$ is the mean $i^{th}$ velocity component. It should be noted here that Wang et al.~\citep{wang2022controlling} have calculated the TKEP on $z=0$ plane using the expression $\text{TKEP} = -\overline{u^{'}_x u^{'}_y} \times \frac{\partial \overline{u_x}}{\partial y}$. This equation is valid only for a two-dimensional plane; however, blood flow in the cardiovascular domain is inherently three-dimensional, and therefore all components of the Reynolds normal and shear stresses, as well as velocity, must be taken into account. This modification has been explicitly made in Eq.~\eqref{eq:TKEP} of this work, and accordingly Fig.~\ref{TKEP} displays the TKEP on $z = 0$ plane. It can be seen that most of the TKEP occurs between the two leaflets, extending somewhat into the sinus region and further downstream, to about two diameters beyond the trailing edges of the leaflets, although with diminishing intensity. However, most of the TKEP is observed in the valve (region B) and sinus (region C), irrespective of the presence or absence of the VGs. In particular, the introduction of VGs on the heart valve leaflets of either type, i.e., rectangular or elliptic, has shown deterioration in the TKEP values, which is illustrated through a histogram plot in Fig.~\ref{VATKEP} by computing the volume-average TKEP (VATKEP) in the valve and sinus region. Mathematically, $\text{VATKEP} = \frac{1}{V_{BC}} \int_{V_{BC}} \text{TKEP}~dV_{BC}$, where $V_{BC}$ is the volume of the region B and C, i.e., valve and the sinus region. Clearly, this plot suggests that EVGs are better suited than RVGs for TKEP reduction, especially when placed near the leading edge of the leaflets; for example, see the difference between SJMVEVG-I and SJMVRVG-I and between SJMVEVG-II and SJMVRVG-II.

\subsection{Pressure distribution}

\begin{figure}
    \centering
    \includegraphics[trim=0cm 0cm 0cm 0cm,clip,width=13cm]{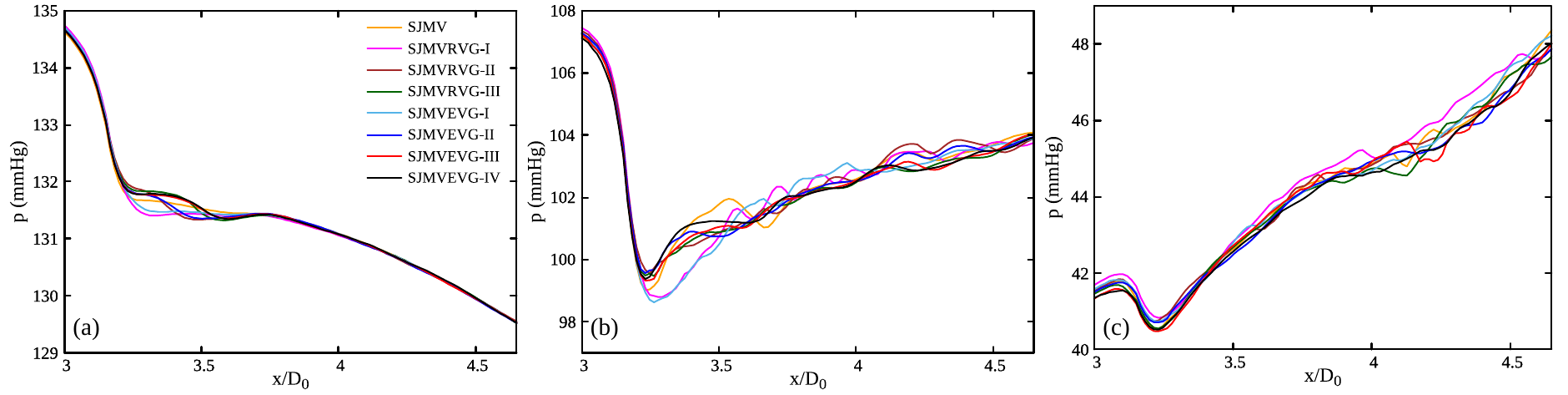}
    \caption{Variation of pressure along the centerline ($x$-axis) in the valve and sinus region of the artery for various configurations presented in Table~\ref{table:HV_design}. (a) time-instant $t = \text{T1} \approx 1965$ ms (mid-acceleration phase), (b) time-instant $t = \text{T2} \approx 2074$ ms (peak systolic phase), and (c) time-instant $t = \text{T3} \approx 2170$ ms (mid-deceleration phase).}
    \label{pCenterline}
\end{figure}

The spatial distribution of pressure along the centerline of the artery ($x$-axis), specifically focused in the valve and sinus region, is illustrated in Fig.~\ref{pCenterline} at three distinct stages of the cardiac cycle (T1, T2, and T3) for all the configurations detailed in Table~\ref{table:HV_design}. During the mid-acceleration phase ($t = \text{T1}$), sub-Fig.~\ref{pCenterline}(a), the upstream pressure is consistently higher for all configurations (with or without VGs) due to the relatively lower magnitude of blood flow near the heart valve leaflets, in accordance with Bernoulli's principle. As blood moves further along the centerline, the pressure decreases, followed by a partial recovery further downstream of the heart valve region. This drop in pressure is again supported by Bernoulli's principle, where blood flow is accelerated in the central region in between the heart valve leaflets (see the magnitude of the central jet in Fig.~\ref{Umagt1}) primarily due to the reduction in the cross-sectional area. Notably, at this time instant of the cardiac cycle, a very minute difference is observed in the pressure values, especially in the regions proximal to the location of the VGs. Likewise the T1 stage, the trend in pressure is similar at the T2 stage (see sub-Fig.~\ref{pCenterline}(b)), although the pressure values in the upstream region decrease due to the increase in blood flow strength. Not only that, the pressure variation further indicates that blood flow is not smoothly varying in the cardiovascular domain; instead, it exhibits unstable or turbulent-like features, as evidenced by the disrupted profiles. In contrast, during the mid-deceleration phase ($t = \text{T3}$) of the cardiac cycle, a retrograde pressure gradient exists, with downstream pressure higher than upstream, due to rapid deceleration of blood flow, ultimately resulting in backflow. However, once again, the influence of incorporating VGs is seen to be minimal on the pressure distribution at this instant of the pulsating waveform. Altogether, for the combinations listed in Table~\ref{table:HV_design}, the deviation in pressure trend is found to be significantly less, especially in the upstream region of the BMHV.

\begin{figure}
    \centering
    \includegraphics[trim=0cm 0cm 0cm 0cm,clip,width=13cm]{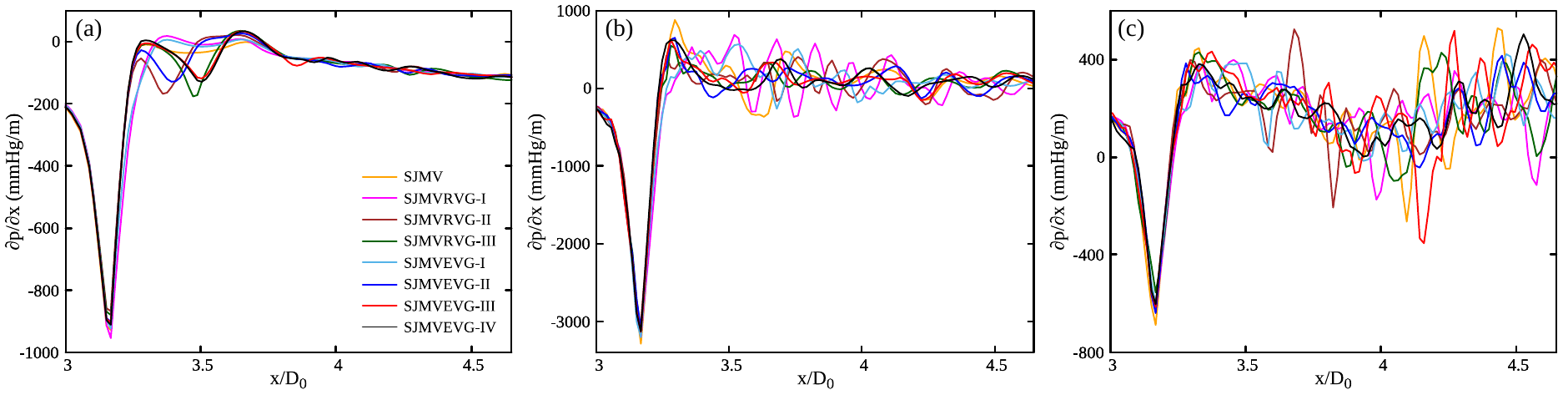}
    \caption{Variation of pressure gradient along the centerline ($x$-axis) in the valve and sinus region of the artery for various configurations presented in Table~\ref{table:HV_design}. (a) time-instant $t = \text{T1} \approx 1965$ ms (mid-acceleration phase), (b) time-instant $t = \text{T2} \approx 2074$ ms (peak systolic phase), and (c) time-instant $t = \text{T3} \approx 2170$ ms (mid-deceleration phase).} 
    \label{gpCenterline}
\end{figure}

To critically and explicitly evaluate the differences between the pressure profiles observed in Fig.~\ref{pCenterline}, the pressure gradients are computed inside the computational domain and presented along the centerline in a similar fashion to that presented earlier in Fig.~\ref{gpCenterline}. From this plot, it can be corroborated that the steepest pressure gradient at three distinct stages of the cardiac cycle, namely, T1 (sub-Fig.~\ref{gpCenterline}(a)), T2 (sub-Fig.~\ref{gpCenterline}(b)), and T3 (sub-Fig.~\ref{gpCenterline}(c)) occurs around $x \approx 3.2 D_0$, irrespective of the presence or absence of the VGs. At the mid-acceleration stage, the variation in the pressure gradient is minimal across different VG configurations on the valve leaflets. However, as the time progresses in the cardiac cycle, differences become noticeable, particularly in the downstream sinus region. Maximum differences are observed at the mid-deceleration stage, where an adverse gradient exists. For certain VG configurations, the pressure gradient shows less variation than in the case without VGs; for instance, see the SJMVEVG-IV case. This could be a benefit of this configuration, as less variation in the pressure gradient could cause less damage to blood cells.

\subsection{Clinical indices}

\begin{figure}
    \centering
    \includegraphics[trim=0cm 0cm 0cm 0cm,clip,width=8cm]{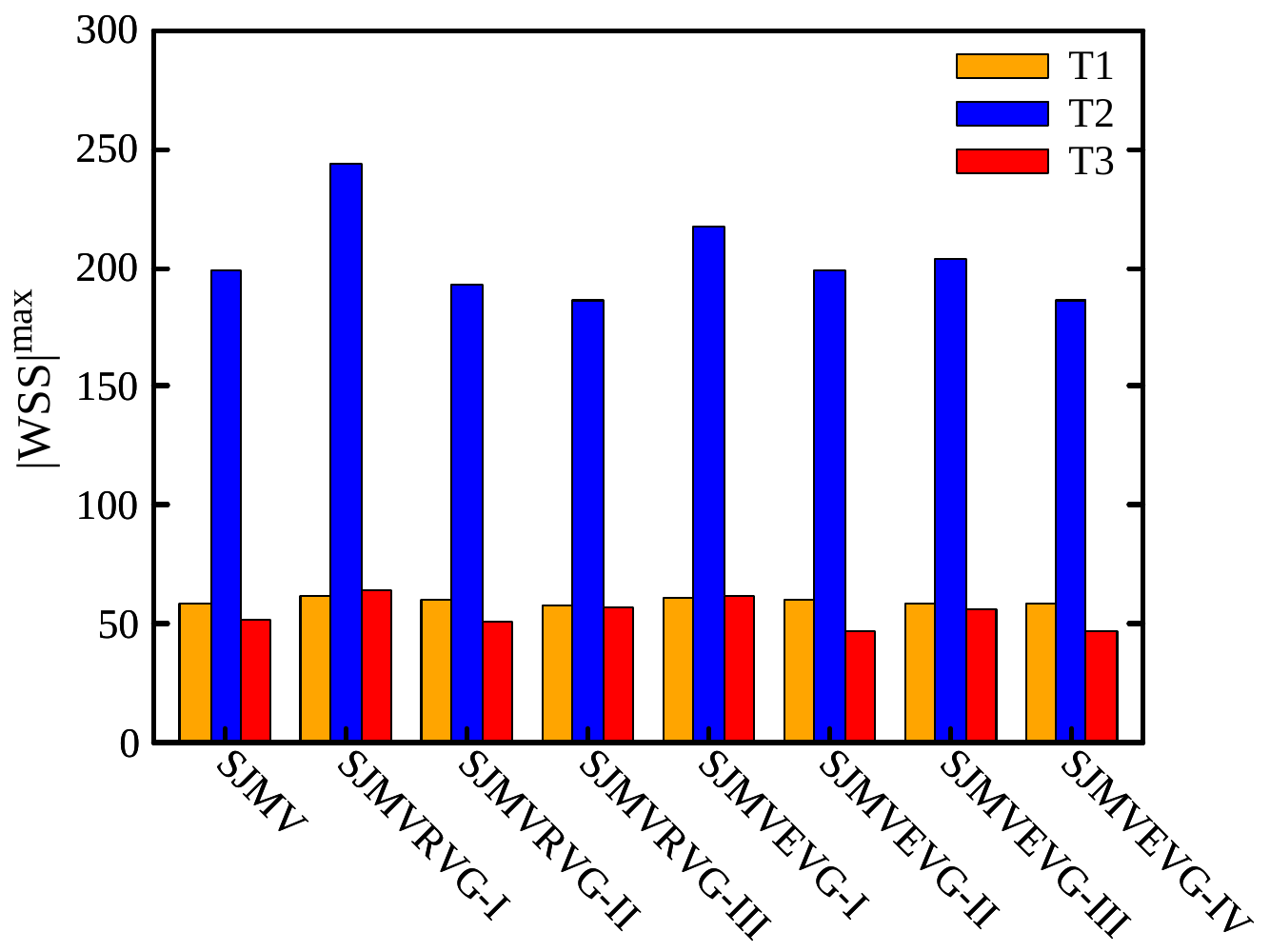}
    \caption{Histogram plot showing the maximum WSS calculated on the artery walls during $t = \text{T1} \approx 1965$ ms (mid-acceleration), $t = \text{T2} \approx 2074$ ms (peak systolic), and $t = \text{T3} \approx 2170$ ms (mid-deceleration) stage of the cardiac cycle for various configurations presented in Table~\ref{table:HV_design}.} 
    \label{WSSmax}
\end{figure}

\begin{figure}
    \centering
    \includegraphics[trim=0cm 0cm 0cm 0cm,clip,width=8cm]{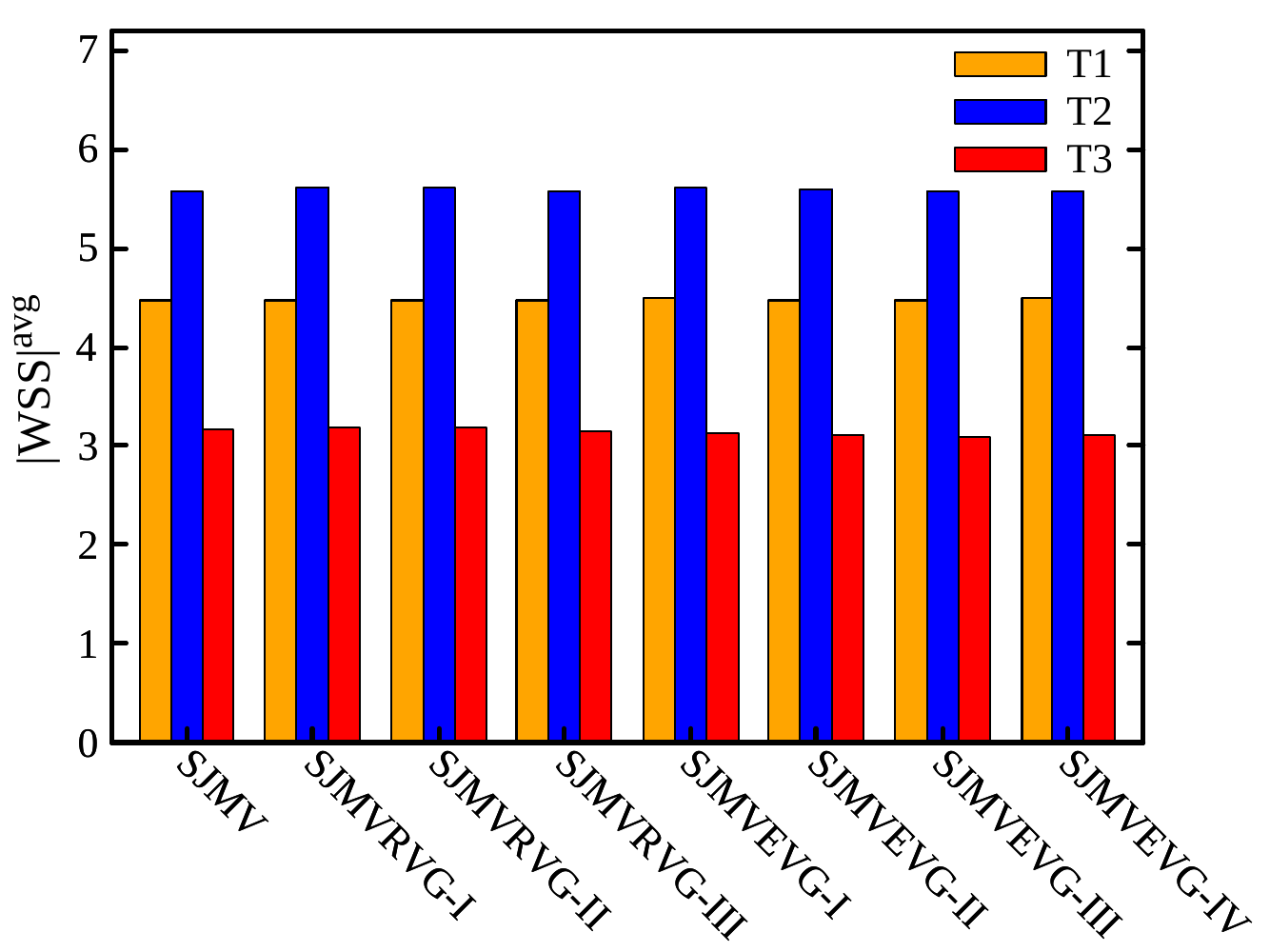}
    \caption{Histogram plot showing the surface-averaged WSS calculated on the artery walls during $t = \text{T1} \approx 1965$ ms (mid-acceleration), $t = \text{T2} \approx 2074$ ms (peak systolic), and $t = \text{T3} \approx 2170$ ms (mid-deceleration) stage of the cardiac cycle for various configurations presented in Table~\ref{table:HV_design}.} 
    \label{WSSavg}
\end{figure}

From the perspective of holding greater clinical relevance to the surface-averaged metrics, in this sub-section, we first portray the histogram plots for the maximum and surface-averaged wall shear stress for all the combinations of VGs simulated in the presented study across three different time instants of the cardiac cycle (T1, T2, and T3), in Figs.~\ref{WSSmax} and~\ref{WSSavg}, respectively. The expression used to evaluate the WSS on the artery walls is given by: $\text{WSS} = \eta |\bm{\dot{\gamma}_w}|$, where $|\bm{\dot{\gamma}_w}|$ is the tangential shear-rate acting on the surface of the artery wall. Coming to the trend, the maximum WSS values in Fig.~\ref{WSSmax} are highest during the time instant T2 among the three time-instants (T1, T2 and T3) and are sensitive to the VG type, i.e., whether it is rectangular or elliptic, as well as their location from the leading edge of the heart valve leaflets. In particular, at time instant T1 of the cardiac waveform, the maximum WSS values are mostly similar or increase by at most $5\%$ in the presence of VGs, as compared to the standard SJMV case. However, during the systolic peak of the cardiac cycle ($t = \text{T2}$), the role of VGs becomes a bit complex. For instance, when either of the RVGs or EVGs is placed close to the leading edge of the leaflets (e.g, see the configurations SJMVRVG-I and SJMVEVG-I), the maximum values of WSS are seen to be higher than in the standard SJMV case by approximately $22.71\%$ and $9.25\%$, respectively. As the distance of VGs increases from the leading edge of the two leaflets, RVGs are found to reduce the maximum WSS values by around $2.70\%$ (SJMVRVG-II) and $6.30\%$ (SJMVRVG-III), whereas EVGs have minimal influence. Interestingly, for the SJMVEVG-IV arrangement, the maximum WSS values reduce by around $5.93\%$ at this peak phase of the cardiac cycle. During the mid-deceleration stage ($t = \text{T3}$), the configurations SJMVRVG-I and SJMVEVG-I again show a floundering performance, with maximum WSS values deviating by more than $18\%$ from the standard SJMV case. Although other arrangements, such as SJMVRVG-III and SJMVEVG-III, also display poor performance (divergence $> 6.87\%$), configurations such as SJMVEVG-II and SJMVEVG-IV have reduced the maximum WSS values from the standard SJMV case by more than $9.42\%$.

Despite noticing a discrepancy in the maximum WSS values, the surface-averaged WSS values are hardly affected by the presence or absence of RVGs or EVGs at any time instant of the cardiac cycle, as shown in Fig.~\ref{WSSavg}. Unlike the occurrence of the highest value of maximum WSS during T2, the highest value of surface-averaged WSS also occurs at T2; however, the lowest value (among T1, T2 and T3) of surface-averaged WSS takes place during the T3 stage of the cardiac cycle, with the WSS magnitude being around $30\%$ smaller than that seen at T1. In contrast, the maximum WSS values were comparable during T1 and T3, at least in some configurations with the presence of VGs; for example, see the configurations SJMVRVG-I, SJMVRVG-III, SJMVEVG-I, and SJMVEVG-III in Fig.~\ref{WSSmax}. Note that the negligible changes evidenced in the surface-averaged values of WSS in the present study at different time instants were also previously seen in our earlier work, wherein the impact of partial and/or free-slip conditions was employed on the heart valve leaflets to explicitly see the influence of superhydrophobicity~\citep{chauhan2026impact}.

\begin{figure}
    \centering
    \includegraphics[trim=0cm 0cm 0cm 0cm,clip,width=8cm]{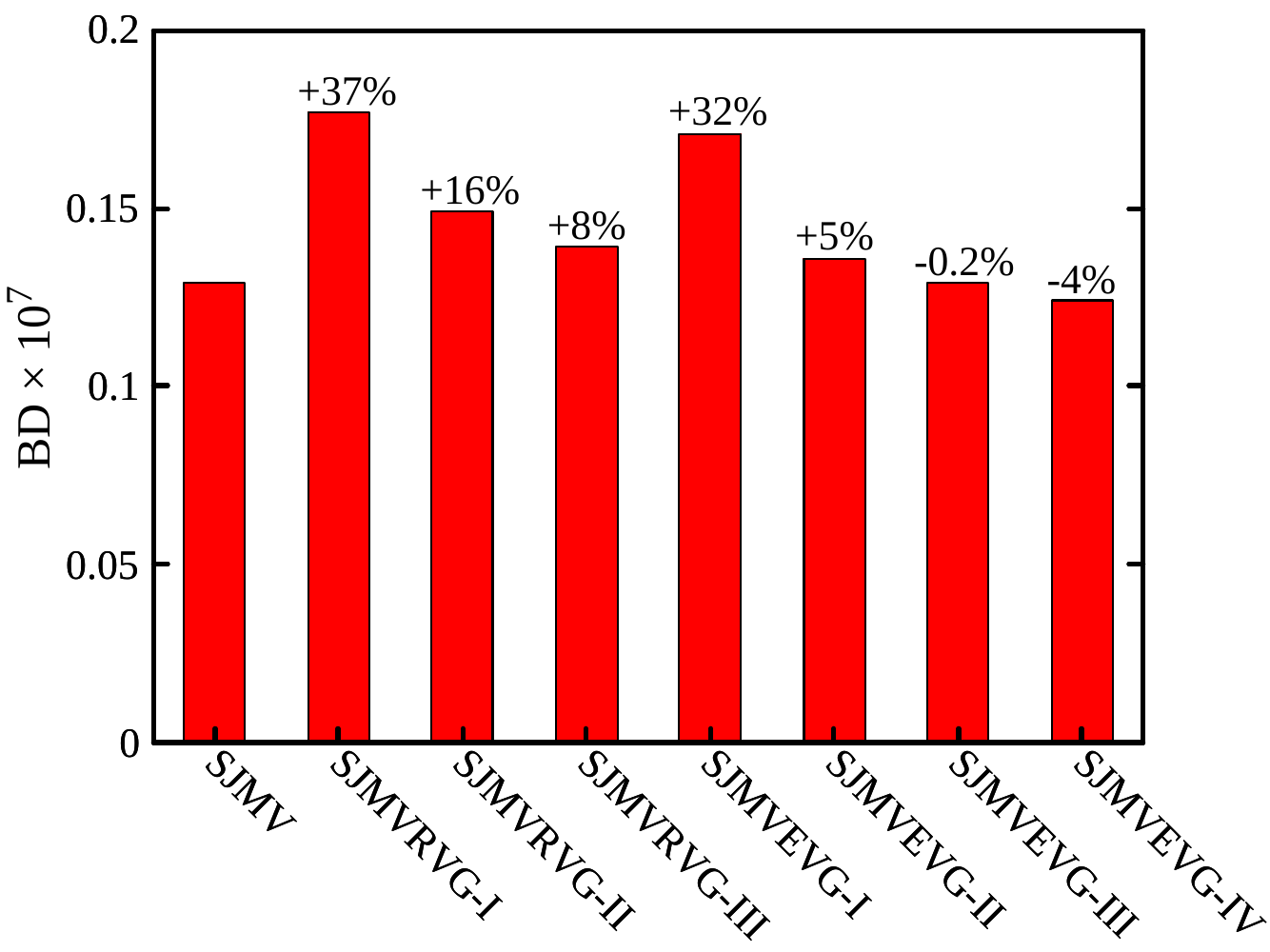}
    \caption{Histogram plot showing the average BD during the entire third cycle for various configurations presented in Table~\ref{table:HV_design}. Note that BD is calculated here only in the valve and sinus region, i.e., region B and C.} 
    \label{BD}
\end{figure}

Next, we compute the blood damage (BD) parameter from the calculated velocity and stress fields within the cardiovascular domain. BD is an important parameter that is connected to shear-induced trauma and hemolysis. In the present study, we evaluate BD using the idea proposed by Garon and Farinas~\citep{garon2004fast}:

\begin{equation}
    \text{BD} = \left(\overline{D_{I}}\right)^{0.785}
\end{equation}
Where $\overline{D_{I}}$ refers to the average linear blood damage given by~\citep{garon2004fast}:

\begin{equation}
    \overline{D_{I}} = \frac{1}{Q_0} \int_{V_{BC}} \sigma~dV_{BC}
\end{equation}

\begin{equation}
    \sigma = (3.62 \times 10^{-7})^{1/0.785} (\tau_{vm})^{2.416/0.785}
\end{equation}
Where $Q_0$ is the average volumetric flow rate throughout the cardiac cycle, $\sigma$ is the hemolysis production rate per unit time, and $\tau_{vm}$ is the von Mises stress calculated using the following expression~\citep{han2022models}:

\begin{equation}
    \tau_{vm} = \left[\frac{1}{6}\left\{ (\tau_{xx} - \tau_{yy})^2 + (\tau_{yy} - \tau_{zz})^2 + (\tau_{zz} - \tau_{xx})^2 \right\} + \tau^2_{xy} + \tau^2_{yz} + \tau^2_{xz}\right]^{0.5}
\end{equation}

Figure~\ref{BD} demarcates the histogram plot of the average BD parameter during the entire third cardiac cycle for different combinations listed in Table~\ref{table:HV_design}. Evidently, in the present study, the inclusion of RVGs on heart valve leaflets has been shown to degrade the BD parameter compared to the standard SJMV case under otherwise identical conditions. For instance, the disparity is calculated to be about $37\%$, $16\%$, and $8\%$ for SJMVRVG-I, SJMVRVG-II, and SJMVRVG-III, respectively. On the other hand, the influence of EVGs on the BD parameter is sensitive to their location and consistently lower than that of RVGs under similar configurations. In particular, for the SJMVEVG-I and SJMVEVG-II arrangements, the BD values are approximately $32\%$ and $5\%$ higher than in the standard SJMV case; however, for SJMVEVG-III, they remain nearly the same as in SJMV. Interestingly, in the SJMVEVG-IV configuration, a decrease in BD of about $4\%$ is observed. This particular arrangement of EVGs (SJMVEVG-IV) on the heart valve leaflets has also shown a decline in the TKEP ($\sim 4.4\%$, Fig.~\ref{VATKEP}) and maximum WSS values ($\sim 5.93\%$ and $\sim 9.42\%$ during systolic peak and mid-deceleration phase, respectively, Fig.~\ref{WSSmax}). Therefore, in the combinations of VGs (RVGs or EVGs) placed on the BMHV that we tested, we have found that SJMVEVG-IV has better overall performance than other arrangements.

\subsection{Dynamic mode decomposition analysis}

The dynamic mode decomposition (DMD) analysis is a popular data-driven dimensionality reduction technique, originally proposed by Schmid~\citep{schmid2011application}. With a sequence of temporally equispaced snapshots of a particular flow field, either in terms of velocity or vorticity, DMD computes a set of modes that are very useful for extracting the dominant coherent structures and their temporal dynamics from the given complex data set. Although DMD is often utilised in fluid dynamics, its use in cardiovascular flows has been under-exploited~\citep{di2019reduced}. Therefore, in the present study, we further delve into the complex blood flow dynamics, particularly the third cardiac pulse of the implemented pulsatile waveform, across all BMHV cases with VGs listed in Table~\ref{table:HV_design}. Specifically, we collect about 370 snapshots of the out-of-plane vorticity field ($\omega_z$) on the $z = 0$ plane only during the systolic stage of the cardiac cycle, sampled at a uniform temporal interval of 1 ms, arranged in the following form

\begin{equation}
S_1 = \begin{bmatrix} 
| & | & & | \\
s_{1} & s_{2} & \cdots & s_{m-1} \\
| & | & & | 
\end{bmatrix}
\quad
S_2 = \begin{bmatrix} 
| & | & & | \\
s_{2} & s_{3} & \cdots & s_{m} \\
| & | & & | 
\end{bmatrix}
\end{equation}
where $s_{j} \in \mathbb{R}^{n}$ represents the vorticity field at time $t_j$.  
Since DMD approximate the temporal evaluation of consecutive snapshots via a linear operator $\bm{P}$, it can be expressed as
\begin{equation}
    S_2 \approx \bm{P}S_1     
\end{equation}
Owing to the fact that $\bm{P}$ is high-dimensional in nature, in general, a reduced-order approximation $\bm{A}$ of $\bm{P}$ is evaluated, which is typically based on minimising the residual $\mathbf{r} = S_{2} - \bm{A}$ $S_{1}$ in a least-squares sense using singular value decomposition (SVD). This results in a low-dimensional representation of the dominant flow dynamics. Moreover, the eigendecomposition of $\bm{A}$ yields the DMD eigenvalues (Ritz values, $\lambda_j$) and eigenvectors, which are then used to reconstruct the associated DMD modes ($\phi_j$) that correspond to the coherent spatial structures of the flow. To relatively quantify the importance of each DMD mode, its amplitude $b_j$ is computed by projecting the initial snapshot onto the DMD modal basis as $\bm{b} = \bm{\phi^{\dagger}} s_1$, where $\bm{\phi^{\dagger}}$ is the Moore-Penrose pseudoinverse of $\bm{\phi}$. The magnitude of $b_j$ typically quantifies the energy contribution of each DMD mode, thereby enabling identification of the dominant coherent structures governing the blood flow dynamics in the present cardiovascular domain. 

\begin{figure}
    \centering
    \includegraphics[trim=0cm 0cm 0cm 0cm,clip,width=13cm]{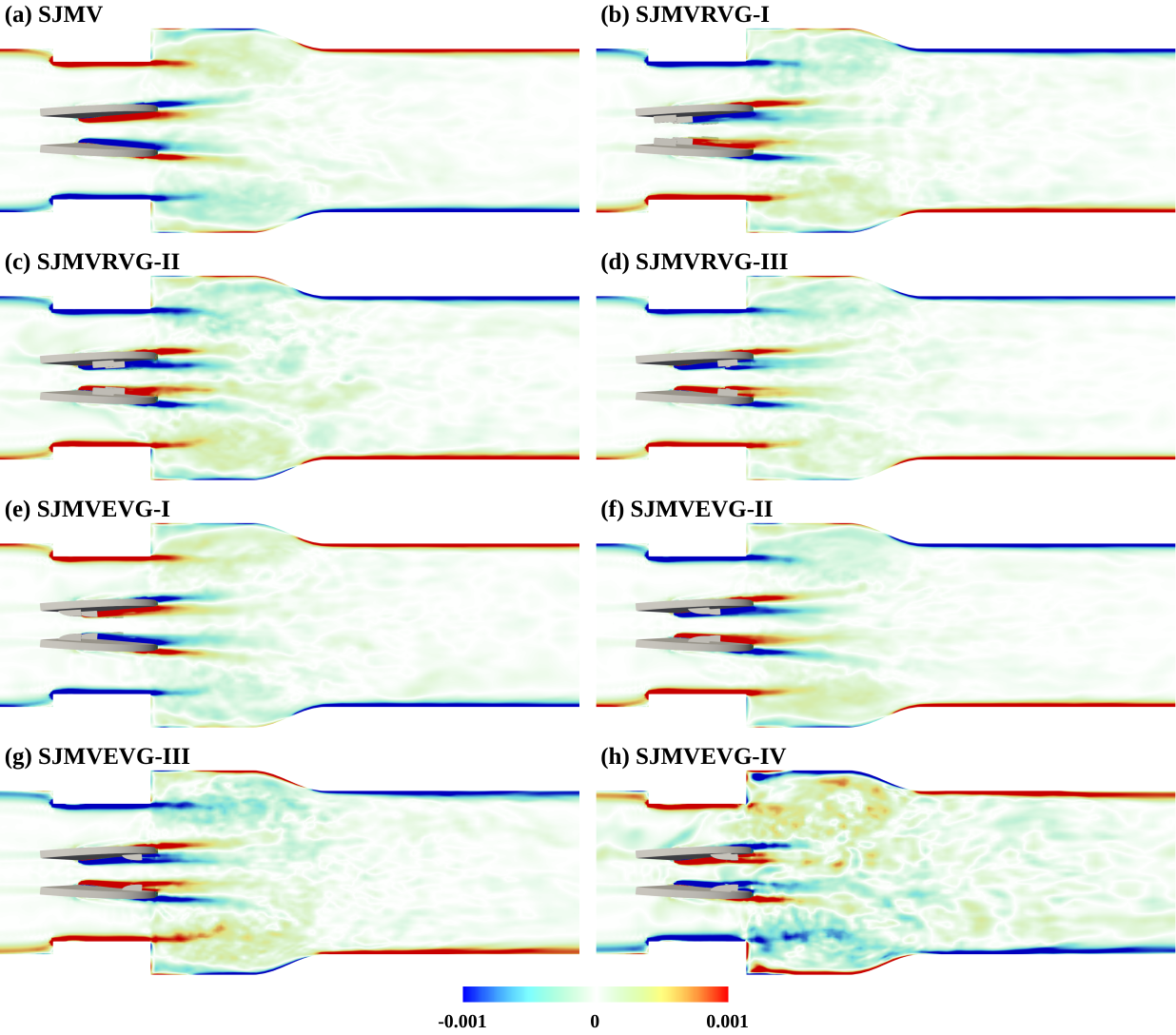}
    \caption{Visual representation of mean DMD mode for various configurations presented in Table~\ref{table:HV_design}.} 
    \label{meanMode}
\end{figure}

\begin{figure}
    \centering
    \includegraphics[trim=0cm 0cm 0cm 0cm,clip,width=13cm]{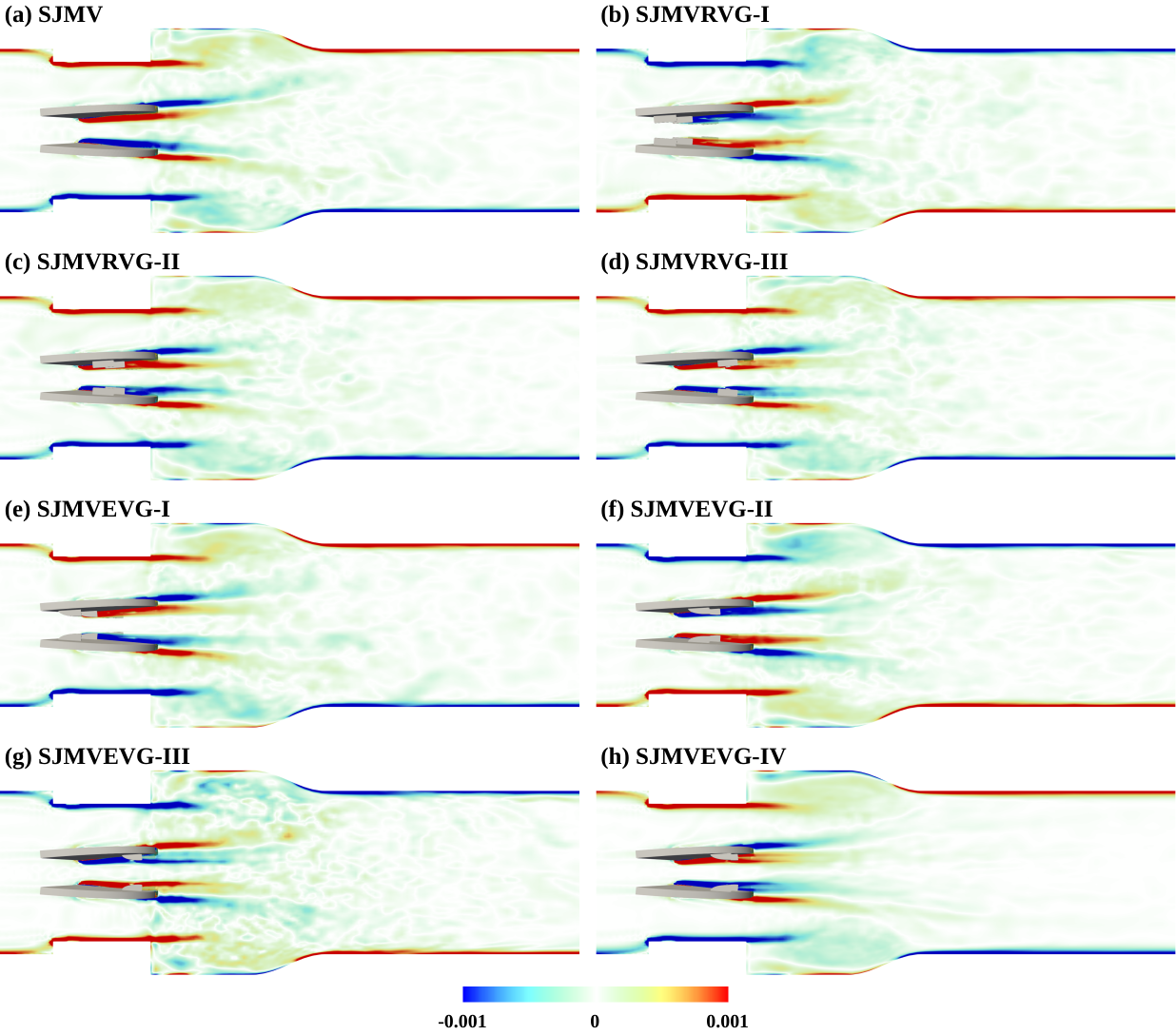}
    \caption{Visual representation of DMD mode-1 for various configurations presented in Table~\ref{table:HV_design}.} 
    \label{mode1}
\end{figure}

\begin{figure}
    \centering
    \includegraphics[trim=0cm 0cm 0cm 0cm,clip,width=13cm]{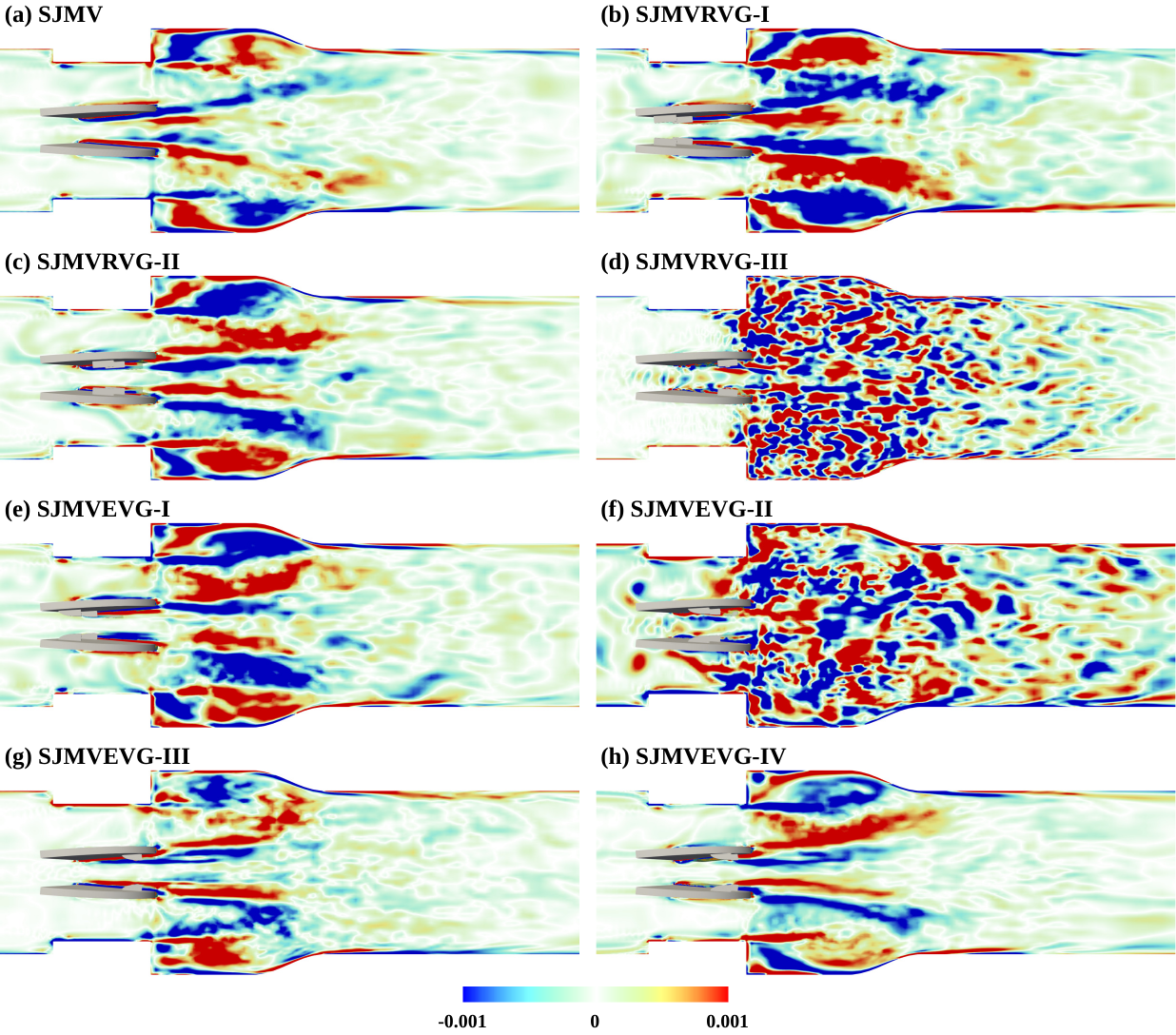}
    \caption{Visual representation of DMD mode-2 for various configurations presented in Table~\ref{table:HV_design}.} 
    \label{mode2}
\end{figure}

\begin{figure}
    \centering
    \includegraphics[trim=0cm 0cm 0cm 0cm,clip,width=13cm]{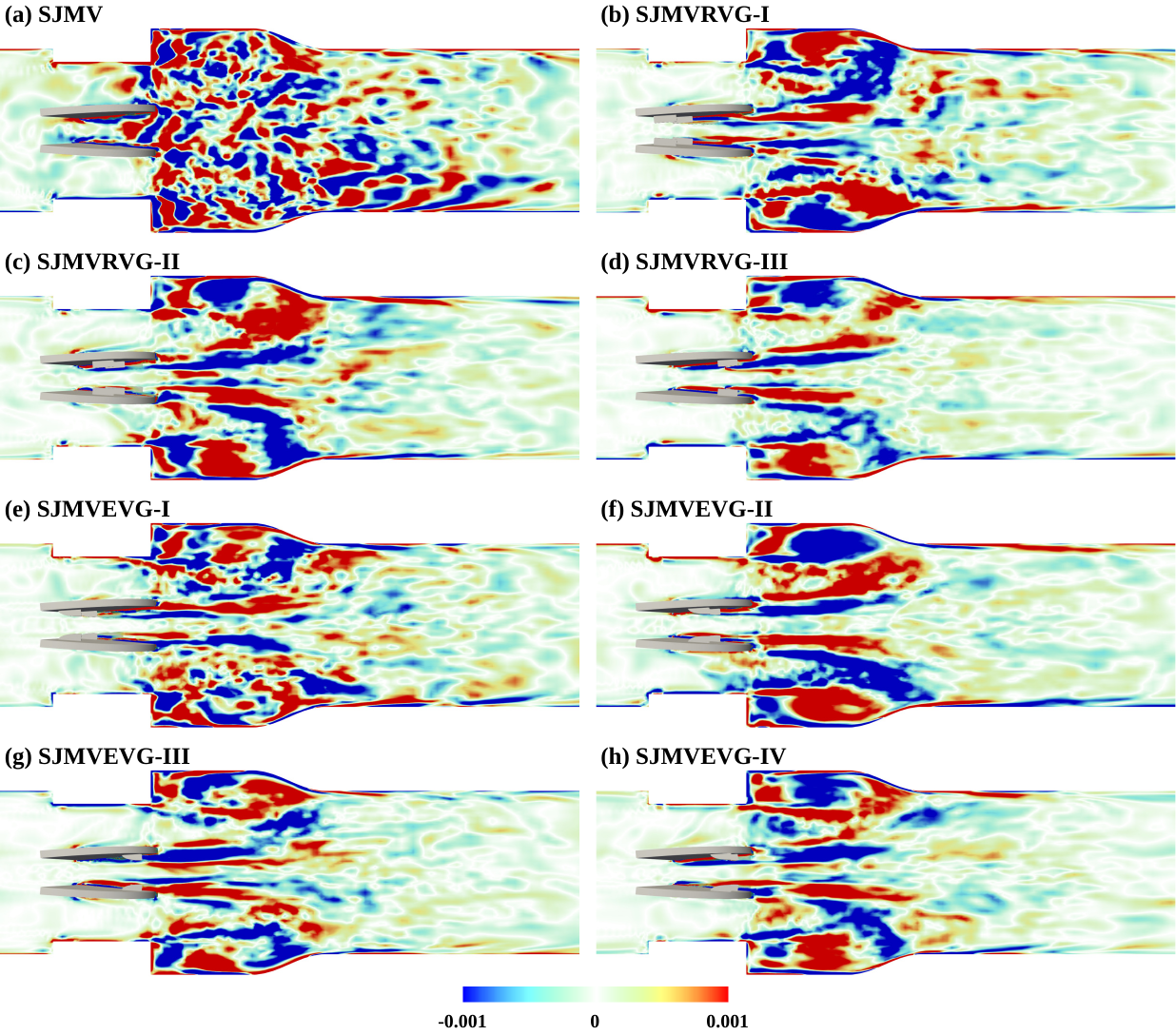}
    \caption{Visual representation of DMD mode-3 for various configurations presented in Table~\ref{table:HV_design}.} 
    \label{mode3}
\end{figure}

The DMD analysis demarcating the mean and the first three modes is shown in Figs.~\ref{meanMode}-\ref{mode3}, respectively, for all BMHV configurations with VGs listed in Table~\ref{table:HV_design}. For the mean DMD mode (Fig.~\ref{meanMode}), regardless of the presence or absence of the VGs, a concentrated coherent flow structure is formed at the leading edge of the valve leaflets, gradually developing along both the leaflet surfaces and extending downstream. A similar concentrated coherent flow structure is also formed along the wall of the artery, especially in the valve region and aortic side chamber, although with a diminished thickness compared to the one near the leaflets. For the standard SJMV case (sub-Fig.~\ref{meanMode}(a)), the structures near the trailing edge of the leaflets seem to be stable. However, with the introduction of VGs, the length and/or width of these structures is seen to increase/decrease depending on the location of the VGs from the leading edge (see sub-Figs.~\ref{meanMode}(b)-(h)). Interestingly, for the SJMVEVG-IV case (sub-Fig.~\ref{meanMode}(h)), the mean mode indicates a distinct behaviour, as the structures, particularly originating from the sides facing the other leaflet, are evidenced to be relatively smaller and unstable than those in the SJMV case. Not only that, but small-scale structures are also observed in the sinus region for the SJMVEVG-IV configuration, which are absent in the SJMV case. This suggests that the introduction of EVGs near the trailing edge of leaflet fragments large structures into smaller ones, potentially affecting the intricate hemodynamics and damage to blood cells during the systolic phase of the cardiac cycle, as also seen before.      

Next, the representation for DMD mode-1 is illustrated in Fig.~\ref{mode1}, which captures the von Kármán vortex-shedding patterns in the wake of the heart valve leaflets. The associated frequency with this mode ($St_1 = \omega_j/2\pi$), where $\omega_j = ln|\lambda_j|/\Delta t$ represents the rate of modal growth or decay, is also calculated for all the configurations. For instance, the values of $St_1$ for mode-1 are obtained as 2.590, 2.319, 2.106, 2.504, 2.380, 2.513, 2.727, and 1.728 for SJMV, SJMVRVG-I, SJMVRVG-II, SJMVRVG-III, SJMVEVG-I, SJMVEVG-II, SJMVEVG-III, and SJMVEVG-IV arrangements, respectively. This clearly indicates that incorporating VGs on the heart valve leaflets generally lowers the vortex-shedding frequency. Interestingly, the configuration SJMVEVG-IV lowers this vortex-shedding frequency the most, by about $33\%$, which has the tendency to decrease the maximum WSS and blood damage as evidenced before in Figs.~\ref{WSSmax} and \ref{BD}, respectively. Coming to the trend of Fig.~\ref{mode1}, for most of the cases, the coherent structures at this mode are attached to the leaflet surfaces, which are slightly distorted, mostly extending into the sinus region. For the standard SJMV (sub-Fig.~\ref{mode1}(a)), these structures, particularly the ones between the artery wall and leaflet surface, extend the most compared to the configurations with VGs, sub-Figs.~\ref{mode1}(b)-(h). For the SJMVEVG-IV setup, the vortex structures are shown to be more stable and less extended in the sinus region, indicating that blood flow is overall suppressed amid the chaos of vortex dynamics past the heart valve's bileaflet structures. This again highlights that the SJMVEVG-IV configuration has better overall performance than the other VG configurations and the standard SJMV case.  

The higher modes, i.e., mode-2 and mode-3 are shown in Figs.~\ref{mode2} and \ref{mode3}, respectively. These two DMD modes further provide clear evidence of intricate blood flow structures, extracting finer details and the noise that evolves within the cardiovascular domain due to the imposition of the complex cardiac cycle. Once again, it can be explicitly seen either in the large sets of small-scale fragment vortex streets or in the increased chaotic vortex structures for the standard SJMV, along with different combinations of RVGs and EVGs, except the SJMVEVG-IV arrangement, wherein such phenomena are comparatively suppressed. All in all, the visualisation of these higher DMD modes again suggests that SJMVEVG-IV is more efficient than other VG configurations.         

\begin{figure}
    \centering
    \includegraphics[trim=0cm 0cm 0cm 0cm,clip,width=13cm]{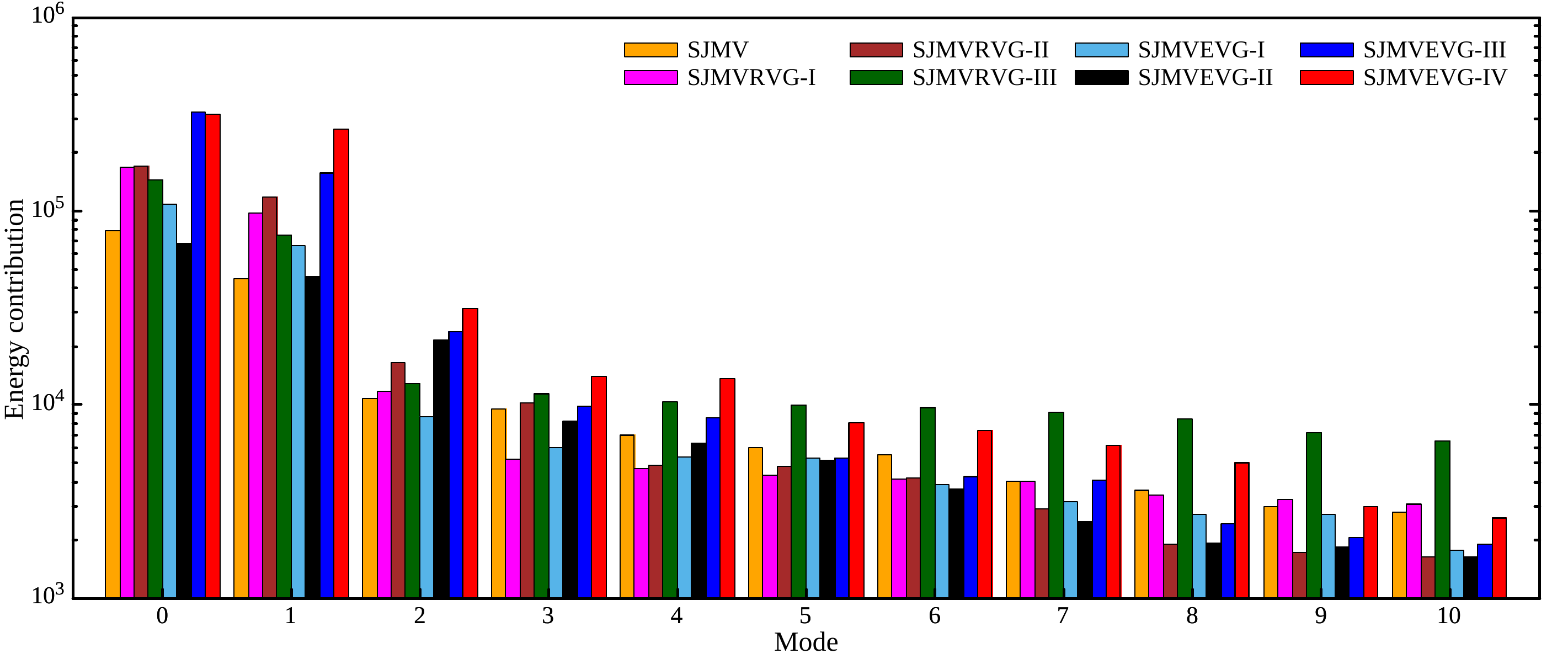}
    \caption{Histogram plot showing the energy contribution of the first few DMD modes for various configurations presented in Table~\ref{table:HV_design}. Here, $0^{\text{th}}$ mode refers to the mean DMD mode.} 
    \label{energy_DMD}
\end{figure}

To further delve deeper and provide more insights from the perspective of DMD analysis, we evaluate the energy contribution of the mean and the first few (10) DMD modes, shown in Fig.~\ref{energy_DMD}. From this histogram plot, it is clear that for all the configurations listed in Table~\ref{table:HV_design}, the energy is mostly clustered towards the mean mode and mode-1. This suggests that lower-order DMD modes (mode-0 and mode-1) contain the majority of the energy of the systolic cardiac signal, and this energy contribution drops precipitously for the higher DMD modes ($\geq 2$), which typically represent the finer details or the noise (due to the rapid separation of the boundary layers), as also pointed out earlier. Moreover, the energy contributions from the EVGs when placed near the trailing edges of the leaflets are highest. Specifically, the SJMVEVG-IV arrangement shows the highest energy contribution, at least for the lower-order DMD modes, indicating the dominant energetic features of the mean mode and mode-1. Therefore, the consequential importance of the SJMVEVG-IV configuration highlighted previously in Figs.~\ref{meanMode} and \ref{mode1} is further supported by this histogram plot.

\section{Conclusions}\label{Conclusions}

In the present study, we performed transient numerical simulations using computational fluid dynamics to examine the influence of two types of VGs, namely elliptic and rectangular, on BMHVs placed at different locations relative to the leading edge of the heart valve leaflets. The physiologically realistic pulsatile flow waveform, which lasts for 860 ms and pertains to 70 beats/min, was incorporated. The results obtained were presented and discussed in terms of velocity magnitude contours, kymograph of the axial velocity field, turbulent kinetic energy production, pressure and its gradient along the centerline, and other clinical indices such as maximum and surface-averaged WSS, blood damage parameter, etc. Our simulations revealed that incorporating VGs (RVGs or EVGs) on the valve leaflet surface affects overall hemodynamics, which is typically sensitive to cardiac waveform timing. For instance, during the mid-acceleration phase of the cardiac cycle ($t = \text{T1}$), subtle changes in the magnitude, length and/or width of the central jet were seen in the presence of VGs. In particular, the velocity magnitude and length in the central region increase in the presence of VGs compared to the standard SJMV case. Furthermore, during the systolic peak ($t = \text{T2}$), the role of VGs has increased significantly compared to the T1 stage. The breakage of a high-magnitude central jet becomes sensitive to the choice of RVGs or EVGs, as well as to their location relative to the leading edge of the leaflets. For example, the central jet breaks into multiple parts when VGs are placed near the leading edge for both RVGs and EVGs; however, when the distance of VGs from the leading edge increases, the breakage happens only for RVGs, but not for EVGs, and as the distance increases, the central jet does not break at all. Additionally, at this stage of the cardiac cycle, the low-magnitude region observed downstream of the trailing edges of the two leaflets increases in length without breakage in the presence of VGs, which is opposite to the standard SJMV case, where breakage occurs. At the mid-deceleration phase ($t = \text{T3}$), the adverse pressure gradient persists, and the overall chaotic behaviour in the cardiovascular domain intensifies, especially in the sinus region and further downstream of it. Interestingly, for some VG configurations, for instance, the SJMVEVG-IV case reduced the velocity-field disruption or broken fragments compared to those seen in the standard SJMV case. 

The kymographs of the axial velocity field revealed that the temporal disruptions were diminished when VGs were introduced, and these disruptions further decreased when the distance of the VGs increased from the leading edge, particularly more evident in EVGs than RVGs. The TKEP, especially in the valve and sinus regions, always deteriorated with the addition of VGs, with the effect more pronounced when VGs were placed near the valve's leading edge. The pressure variation, its gradient along the centerline, and surface-averaged WSS were found to be nearly identical regardless of the time stage of the pulsating waveform and the presence or absence of the VGs. However, the maximum WSS values were observed to increase or decrease depending on the time instant of observation, the location relative to the leading edge of the VGs, and their type, i.e., EVG or RVG. For instance, compared to standard SJMV, the maximum WSS remains almost the same or increases by at most $5\%$ in the presence of VGs during the T1 time instant, its value at the systolic peak ($t = \text{T2}$) and mid-deceleration phase ($t = \text{T3}$) were evidenced to be accentuated by more than $9\%$ and $18\%$, respectively, especially when the VGs are placed close to the leading edge. Interestingly, the SJMVEVG-IV configuration has reduced the maximum WSS values by approximately $6\%$ and $9\%$ during the T2 and T3 stages of the pulsating waveform, respectively. Furthermore, subsuming RVGs at any location has shown a detrimental effect on the blood damage parameter; however, when EVGs are placed close to the trailing edges of the leaflets, especially the SJMVEVG-IV arrangement, the blood damage decreased by about $4\%$. Finally, the data-driven DMD technique applied to the vorticity field during the systolic stage of the cardiac cycle was analysed across all eight configurations to uncover hidden coherent structures in the cardiovascular domain. The energy contributions of the first few DMD modes revealed that the energy is mostly inclined towards the lower-order DMD modes, particularly the mean mode and mode-1. Consequently, distinct vortex structures were observed, particularly in the SJMVEVG-IV configuration compared to the standard SJMV case. For instance, in DMD mode-1, the coherent vortex structures attached to the leaflet surfaces were more stable and less extended in the sinus region than in the SJMV case, indicating suppression of chaotic vortex dynamics. This was further justified by the calculation of the vortex-shedding frequency corresponding to mode-1, wherein SJMVEVG-IV lowers this frequency by about $33\%$ compared to the SJMV setup. Therefore, among the combinations of EVGs and RVGs used in the present work, SJMVEVG-IV has shown better overall performance with respect to improvements in TKEP, maximum WSS, vorticity dynamics, and blood damage, and can be considered a novel vortex generator for the design and development of next-generation BHMVs.

\backmatter

\bmhead{Acknowledgements}

We acknowledge the National Supercomputing Mission (NSM) for providing computing resources of ‘PARAM Smriti’ at NABI, Mohali (accessed by CS) and ‘PARAM Himalaya’ at IIT Mandi (accessed by AC), which are implemented by C-DAC and supported by the Ministry of Electronics and Information Technology (MeitY) and Department of Science and Technology (DST), Government of India. AC sincerely thanks the Ministry of Education, Government of India, for the financial support provided by the PMRF (Cycle-9).

\bmhead{Conflict of Interest}
There is no conflict of interest present in the present manuscript. 

\bibliographystyle{sn-basic}
\bibliography{References}

\begin{thebibliography}{45}
\providecommand{\natexlab}[1]{#1}
\providecommand{\url}[1]{{#1}}
\providecommand{\urlprefix}{URL }
\providecommand{\doi}[1]{\url{https://doi.org/#1}}
\providecommand{\eprint}[2][]{\url{#2}}
 \bibcommenthead

\bibitem[{ope(2019)}]{openfoam}
 (2019) {OpenFOAM v7}. \url{https://openfoam.org/version/7/}

\bibitem[{WHO(2025)}]{WHO}
 (2025) \url{https://www.who.int/news-room/fact-sheets/detail/cardiovascular-diseases-(cvds)}

\bibitem[{Ajmone~Marsan et~al.(2023)Ajmone~Marsan, Delgado, Shah, Pellikka, Bax, Treibel, and Cavalcante}]{ajmone2023valvular}
Ajmone~Marsan N, Delgado V, Shah DJ, et~al (2023) Valvular heart disease: Shifting the focus to the myocardium. European Heart Journal 44(1):28--40

\bibitem[{Alemu and Bluestein(2007)}]{alemu2007flow}
Alemu Y, Bluestein D (2007) Flow-induced platelet activation and damage accumulation in a mechanical heart valve: {N}umerical studies. Artificial Organs 31(9):677--688

\bibitem[{Bark~Jr et~al.(2017)Bark~Jr, Vahabi, Bui, Movafaghi, Moore, Kota, Popat, and Dasi}]{bark2017hemodynamic}
Bark~Jr DL, Vahabi H, Bui H, et~al (2017) Hemodynamic performance and thrombogenic properties of a superhydrophobic bileaflet mechanical heart valve. Annals of Biomedical Engineering 45(2):452--463

\bibitem[{Berry et~al.(2012)Berry, Dyer, Cai, Garside, Ning, Thomas, Greenland, Van~Horn, Tracy, and Lloyd-Jones}]{berry2012lifetime}
Berry JD, Dyer A, Cai X, et~al (2012) Lifetime risks of cardiovascular disease. New England Journal of Medicine 366(4):321--329

\bibitem[{Bluestein et~al.(2000)Bluestein, Rambod, and Gharib}]{bluestein2000vortex}
Bluestein D, Rambod E, Gharib M (2000) Vortex shedding as a mechanism for free emboli formation in mechanical heart valves. Journal of Biomechanical Engineering 122(2):125--134

\bibitem[{Chauhan and Sasmal(2024{\natexlab{a}})}]{chauhan2024hemodynamics}
Chauhan A, Sasmal C (2024{\natexlab{a}}) Hemodynamics past a dysfunctional bileaflet mechanical heart valve. International Journal of Engineering Science 205:104154

\bibitem[{Chauhan and Sasmal(2024{\natexlab{b}})}]{chauhan2024influence}
Chauhan A, Sasmal C (2024{\natexlab{b}}) The influence of non-{N}ewtonian behaviors of blood on the hemodynamics past a bileaflet mechanical heart valve. Physics of Fluids 36(10)

\bibitem[{Chauhan and Sasmal(2026)}]{chauhan2026impact}
Chauhan A, Sasmal C (2026) Impact of superhydrophobic slip surface on modulating the hemodynamics of a bileaflet mechanical heart valve under functional and dysfunctional states. Physics of Fluids 38(02)

\bibitem[{Chhabra and Richardson(2011)}]{Chhabra2011}
Chhabra RP, Richardson JF (2011) Non-{N}ewtonian Flow and Applied Rheology: Engineering Applications. Butterworth-Heinemann

\bibitem[{Coffey et~al.(2016)Coffey, Cairns, and Iung}]{coffey2016modern}
Coffey S, Cairns BJ, Iung B (2016) The modern epidemiology of heart valve disease. Heart 102(1):75--85

\bibitem[{Coffey et~al.(2021)Coffey, Roberts-Thomson, Brown, Carapetis, Chen, Enriquez-Sarano, Z{\"u}hlke, and Prendergast}]{coffey2021global}
Coffey S, Roberts-Thomson R, Brown A, et~al (2021) Global epidemiology of valvular heart disease. Nature Reviews Cardiology 18(12):853--864

\bibitem[{Dasi et~al.(2008)Dasi, Murphy, Glezer, and Yoganathan}]{dasi2008passive}
Dasi LP, Murphy DW, Glezer A, et~al (2008) Passive flow control of bileaflet mechanical heart valve leakage flow. Journal of Biomechanics 41(6):1166--1173

\bibitem[{Dasi et~al.(2009)Dasi, Simon, Sucosky, and Yoganathan}]{dasi2009fluid}
Dasi LP, Simon HA, Sucosky P, et~al (2009) Fluid mechanics of artificial heart valves. Clinical and Experimental Pharmacology and Physiology 36(2):225--237

\bibitem[{DeWall et~al.(2000)DeWall, Qasim, and Carr}]{dewall2000evolution}
DeWall RA, Qasim N, Carr L (2000) Evolution of mechanical heart valves. The Annals of Thoracic Surgery 69(5):1612--1621

\bibitem[{Di~Labbio and Kadem(2019)}]{di2019reduced}
Di~Labbio G, Kadem L (2019) Reduced-order modeling of left ventricular flow subject to aortic valve regurgitation. Physics of Fluids 31(3)

\bibitem[{Evangelista et~al.(2025)Evangelista, Pires, and Nogueira}]{evangelista2025chronological}
Evangelista RAA, Pires ALR, Nogueira BV (2025) A chronological history of heart valve prostheses to offer perspectives of their limitations. Frontiers in Bioengineering and Biotechnology 13:1533421

\bibitem[{Ferziger et~al.(2002)Ferziger, Peri{\'c}, and Street}]{ferziger2002computational}
Ferziger JH, Peri{\'c} M, Street RL (2002) Computational methods for fluid dynamics, vol~3. Springer

\bibitem[{Gaidai et~al.(2023)Gaidai, Cao, and Loginov}]{gaidai2023global}
Gaidai O, Cao Y, Loginov S (2023) Global cardiovascular diseases death rate prediction. Current Problems in Cardiology 48(5):101622

\bibitem[{Garon and Farinas(2004)}]{garon2004fast}
Garon A, Farinas MI (2004) Fast three-dimensional numerical hemolysis approximation. Artificial Organs 28(11):1016--1025

\bibitem[{Ge et~al.(2005)Ge, Leo, Sotiropoulos, and Yoganathan}]{Ge2005}
Ge L, Leo HL, Sotiropoulos F, et~al (2005) Flow in a mechanical bileaflet heart valve at laminar and near-peak systole flow rates: {C}{F}{D} simulations and experiments. Journal of Biomechanical Engineering 127(5):782--797

\bibitem[{Gott et~al.(2003)Gott, Alejo, and Cameron}]{gott2003mechanical}
Gott VL, Alejo DE, Cameron DE (2003) Mechanical heart valves: 50 years of evolution. The Annals of Thoracic Surgery 76(6):S2230--S2239

\bibitem[{Groun et~al.(2022)Groun, Villalba-Orero, Lara-Pezzi, Valero, Garicano-Mena, and Le~Clainche}]{Groun2022}
Groun N, Villalba-Orero M, Lara-Pezzi E, et~al (2022) Higher order dynamic mode decomposition: From fluid dynamics to heart disease analysis. Computers in Biology and Medicine 144:105384

\bibitem[{Han et~al.(2022)Han, Zhang, Griffith, and Wu}]{han2022models}
Han D, Zhang J, Griffith BP, et~al (2022) Models of shear-induced platelet activation and numerical implementation with computational fluid dynamics approaches. Journal of Biomechanical Engineering 144(4):040801

\bibitem[{Hatoum and Dasi(2019)}]{hatoum2019reduction}
Hatoum H, Dasi LP (2019) Reduction of pressure gradient and turbulence using vortex generators in prosthetic heart valves. Annals of Biomedical Engineering 47(1):85--96

\bibitem[{Hatoum et~al.(2020)Hatoum, Vallabhuneni, Kota, Bark, Popat, and Dasi}]{hatoum2020impact}
Hatoum H, Vallabhuneni S, Kota AK, et~al (2020) Impact of superhydrophobicity on the fluid dynamics of a bileaflet mechanical heart valve. Journal of the Mechanical Behavior of Biomedical Materials 110:103895

\bibitem[{Head et~al.(2017)Head, {\c{C}}elik, and Kappetein}]{head2017mechanical}
Head SJ, {\c{C}}elik M, Kappetein AP (2017) Mechanical versus bioprosthetic aortic valve replacement. European Heart Journal 38(28):2183--2191

\bibitem[{Jaffer and Whitlock(2016)}]{jaffer2016mechanical}
Jaffer IH, Whitlock RP (2016) A mechanical heart valve is the best choice. Heart Asia 8(1):62--64

\bibitem[{Jayanarasimhan and Balasubramanian(2025)}]{jayanarasimhan2025overview}
Jayanarasimhan K, Balasubramanian NK (2025) An overview of flow control in aerodynamic surfaces using vortex generators. Physics of Fluids 37(3)

\bibitem[{Jirasek(2005)}]{jirasek2005vortex}
Jirasek A (2005) Vortex-generator model and its application to flow control. Journal of Aircraft 42(6):1486--1491

\bibitem[{Murphy et~al.(2010)Murphy, Dasi, Vukasinovic, Glezer, and Yoganathan}]{murphy2010reduction}
Murphy DW, Dasi LP, Vukasinovic J, et~al (2010) Reduction of procoagulant potential of b-datum leakage jet flow in bileaflet mechanical heart valves via application of vortex generator arrays. Journal of Biomechanical Engineering 132(7):071011

\bibitem[{Nitti et~al.(2022)Nitti, De~Cillis, and De~Tullio}]{nitti2022numerical}
Nitti A, De~Cillis G, De~Tullio M (2022) Numerical investigation of turbulent features past different mechanical aortic valves. Journal of Fluid Mechanics 940:A43

\bibitem[{Pibarot and Dumesnil(2009)}]{pibarot2009prosthetic}
Pibarot P, Dumesnil JG (2009) Prosthetic heart valves: Selection of the optimal prosthesis and long-term management. Circulation 119(7):1034--1048

\bibitem[{Schmid(2010)}]{schmid2010dynamic}
Schmid PJ (2010) Dynamic mode decomposition of numerical and experimental data. Journal of Fluid Mechanics 656:5--28

\bibitem[{Schmid(2011)}]{schmid2011application}
Schmid PJ (2011) Application of the dynamic mode decomposition to experimental data. Experiments in Fluids 50(4):1123--1130

\bibitem[{Schmid(2022)}]{schmid2022dynamic}
Schmid PJ (2022) Dynamic mode decomposition and its variants. Annual Review of Fluid Mechanics 54(1):225--254

\bibitem[{Singh et~al.(2023)Singh, Kachel, Castillero, Xue, Kalfa, Ferrari, and George}]{singh2023polymeric}
Singh SK, Kachel M, Castillero E, et~al (2023) Polymeric prosthetic heart valves: A review of current technologies and future directions. Frontiers in Cardiovascular Medicine 10:1137827

\bibitem[{Squiers et~al.(2023)Squiers, Robinson, Audisio, Ryan, Mack, Rahouma, Cancelli, Kirov, Doenst, Gaudino et~al.}]{squiers2023structural}
Squiers JJ, Robinson NB, Audisio K, et~al (2023) Structural valve degeneration of bioprosthetic aortic valves: {A} network meta-analysis. The Journal of Thoracic and Cardiovascular Surgery 166(1):52--59

\bibitem[{Wang et~al.(2022)Wang, Dasi, and Hatoum}]{wang2022controlling}
Wang Z, Dasi LP, Hatoum H (2022) Controlling the flow separation in heart valves using vortex generators. Annals of Biomedical Engineering 50(8):914--928

\bibitem[{Yoganathan et~al.(2004)Yoganathan, He, and Casey~Jones}]{yoganathan2004fluid}
Yoganathan AP, He Z, Casey~Jones S (2004) Fluid mechanics of heart valves. Annual Review of Biomedical Engineering 6(1):331--362

\bibitem[{Yousefi et~al.(2024)Yousefi, Borna, Shirvan, Wen, and Nouri}]{yousefi2024surface}
Yousefi S, Borna H, Shirvan AR, et~al (2024) Surface modification of mechanical heart valves: A review. European Polymer Journal 205:112726

\bibitem[{Yun et~al.(2014)Yun, Dasi, Aidun, and Yoganathan}]{Yun2014}
Yun BM, Dasi LP, Aidun CK, et~al (2014) Computational modelling of flow through prosthetic heart valves using the entropic lattice-boltzmann method. Journal of Fluid Mechanics 743:170--201

\bibitem[{Zakaria et~al.(2017)Zakaria, Ismail, Tamagawa, Aziz, Wiriadidjaja, Basri, and Ahmad}]{Zakaria2017}
Zakaria MS, Ismail F, Tamagawa M, et~al (2017) Review of numerical methods for simulation of mechanical heart valves and the potential for blood clotting. Medical \& Biological Engineering \& Computing 55(9):1519–1548

\bibitem[{Zhao et~al.(2022)Zhao, Jiang, Feng, Liu, Wang, Shen, Chen, Wang, and Liu}]{zhao2022researches}
Zhao Z, Jiang R, Feng J, et~al (2022) Researches on vortex generators applied to wind turbines: A review. Ocean Engineering 253:111266

\end{thebibliography}

\end{document}